\documentclass[showpacs,preprintnumbers,amsmath,amssymb,onecolumn,12pt]{revtex4-1}
\usepackage{changes}
\usepackage{graphicx}
\usepackage{dcolumn}
\usepackage{bm}
\usepackage{subfigure}
\usepackage{float}
\usepackage{multirow}
\usepackage{booktabs}
\usepackage{amsmath}
\usepackage{cases}
\usepackage{latexsym,amssymb}
\usepackage[mathscr]{eucal}
\usepackage{lineno}
\usepackage[bookmarks=true,
   colorlinks=true,
   linkcolor=blue,
   urlcolor=blue,
   citecolor=blue,
   bookmarks=true,
   hyperindex=true
]{hyperref}

\begin{document}
\allowdisplaybreaks[4]
\title{New black-to-white hole solutions with improved geometry and energy conditions}

\author{Zhong-Wen Feng\textsuperscript{1}}
\altaffiliation{Email: zwfengphy@cwnu.edu.cn}
\author{Yi Ling \textsuperscript{2,3,1}}
\altaffiliation{Email: lingy@ihep.ac.cn}
\author{Xiao-Ning Wu\textsuperscript{4,5}}
\altaffiliation{Email: wuxn@amss.ac.cn}
\author{Qing-Quan Jiang\textsuperscript{1}}
\altaffiliation{Email: qqjiangphys@yeah.net}
\vskip 0.5cm
\affiliation{1 School of Physics and Astronomy, China West Normal University, Nanchong 637009, China
\\ 2 Institute of High Energy Physics, Chinese Academy of Sciences, Beijing 100049, China
\\ 3School of Physics, University of Chinese Academy of Sciences, Beijing 100049, China
\\ 4 Institute of Mathematics, Academy of Mathematics and System Science and Hua Loo-Keng Key Laboratory of Mathematics, Chinese Academy of Sciences, Beijing 100190, China
\\ 5 School of Mathematical Sciences, University of Chinese Academy of Sciences, Beijing 100049, China}

\begin{abstract}
We construct new black-to-white hole solutions which connect the geometry of spacetime at some gluing surface inside the horizon. The continuity of the metric can be guaranteed up to the arbitrary order which is controlled by the power factor $n$. This sort of black-to-white holes is characterized by the sub-Planckian scalar curvature, independent of the mass of black-to-white holes. More importantly, we show that the energy condition is only violated within a small region near the gluing surface.  The geodesics of particles within the region from black hole to white hole is also analyzed. It turns out that the matter falling into the black hole may pass through the center without singularity and come out from the white hole. This scenario provides novel ideas for understanding the information loss paradox in traditional black hole physics.
\end{abstract}

\maketitle
\section{Introduction}
\label{intro}
As one of the predictions of the theory of general relativity (GR), black holes are regarded as enigmatic and captivating entities within the vastness of the universe \cite{chb1}. Their extraordinary gravitational pull and distinctive characteristics have long captivated the attention of scientific community. Recent breakthroughs, including the successful detection of gravitational waves by LIGO \cite{cha1} and the observation of black hole shadows by EHT \cite{cha2,chb2}, have unequivocally affirmed the existence of these intriguing objects in the universe. Despite these remarkable achievements, one still needs to face the fact that GR has its own limitations, particularly concerning the issue of singularity \cite{cha3,cha4,chb4,chb4+}. It is widely recognized that classical black holes exhibit a fundamental flaw in the form of UV (ultraviolet) incompleteness. This implies that gravitational collapse inevitably leads to the formation of singularity, a space time region with infinitely large curvature where all matter and energy become concentrated. The presence of singularity profoundly impacts the properties of black holes. Firstly, the singularity represents the culmination of spacetime, where an immense concentration of matter and energy occurs, resulting in an extreme curvature of the surrounding spacetime. This extraordinary condition defies the laws of physics and undermines the coherence of the spacetime fabric. Secondly, when taking the quantum effects into account, the black hole exhibits abundant thermodynamic behavior and loses its energy by Hawking radiation, and inevitably disappears. This evaporation process of black holes brings out more intricate predicaments: the puzzled of ``information loss paradox" and the dilemma of facing the dangerous ``naked singularity" \cite{cha5}.

In order to overcome the difficulties caused by singularity, people are committed to constructing singularity-free black hole solutions, which are known as regular black holes (RBHs) \cite{chb5,chb5+,cha6,cha7}. Within RBHs, curvature invariants exhibit a remarkable behavior, namely remaining finite throughout the spacetime, particularly in the vicinity of the core of black hole. Early attempts at constructing RBHs were made by Sakharov and Gliner \cite{cha8,cha9}, who proposed a novel way to circumvent spacetime singularity by introducing a de Sitter core in place of the vacuum. Subsequently, Bardeen pioneered the development of a valid model for RBHs, known as the Bardeen black hole, by incorporating an $r$-dependent mass function $M(r)$ in lieu of the Schwarzschild black hole mass $M$ \cite{cha10}. This model has garnered significant attention, as it not only successfully evades the singularity predicament but has also been demonstrated to be a solution of Einstein's equation within the framework of nonlinear electrodynamics. Motivated by this approach, a number of RBHs have been constructed, see e.g.,~Refs.~\cite{cha11,cha12,cha13,cha14,chb14,cha15,chc15,chc16,chc17,chc18,chc19,chc20,chc21,chc22,chc23,chc24,chc25,chf1,chf2,chf3,chf4,chf5,chf6,chf7,chf8} and the references therein.

Beyond the conventional approach of substituting the vacuum with alternative forms of matter, the construction of RBHs has also been explored within the framework of modified gravity theories, such as $f(R)$ theory \cite{cha16}, $f(T)$ theory \cite{cha17}, Lovelock gravity \cite{cha18}, and non-commutative geometry \cite{cha18+}. In particular, it has been suggested by several studies \cite{cha19,cha19+,cha20,cha21} that quantum gravity (QG) plays a pivotal role in the formation of RBHs, since QG modifies the curvature of spacetime near the Planck scale, effectively circumventing curvature divergence at the singularity. The investigation of RBHs and their connection to QG is driven by multiple reasons. (i) the RBH models serve as a platform to explore the influence of QG effects in highly gravitational environments, offering insights into spacetime structure, energy density, and related physical properties within a quantum framework; (ii) by examining the properties and characteristics of RBHs, it becomes possible to propose observable signatures and effects that can guide future observations in validating or ruling out various candidates for QG theories and models; (iii) the QG effects in RBHs present a promising avenue for addressing the information loss paradox and the problem of bare singularity in black holes. Motivated by these considerations, numerous RBHs incorporating QG effects have been proposed~\cite{cha4,chr1,chr2,chr3,chr4,cha22,cha23,cha25,cha26,cha27,chb27,chb19,chb20,chy1}.

An interesting aspect of unifying QG with RBHs is the possibility of obtaining black-to-white hole solutions (For recent review we refer to Ref.~\cite{chy3}). Classical relativity dictates that black holes are the ultimate outcome of stellar evolution. However, with the consideration of QG effects, the tendency of black holes to undergo total collapse halts near the Planck scale. This intriguing phenomenon results in the emergence of a spacetime channel at the center of the black hole, facilitating the passage of matter that was absorbed by the black hole. Consequently, this matter is then ejected into a spacetime possessing properties opposite to those of the black hole, known as white holes. The black-to-white hole solutions challenge the conventional notion that black holes are the final stage of stellar evolution. Instead, they demonstrate that matter entering black holes can be subsequently released through white holes, effectively preserving the conservation of information. This intriguing discovery has led researchers to shift their focus towards investigating black-to-white hole solutions and exploring their associated properties, see e.g., Refs.~\cite{chr25,chr26,chr27,chr28,chr29,chr30,chr31,chr32,chr33,chr34,chr35,chr36,chr37,chr38}.

As an illustration, we highlight two notable examples of RBHs with the effects of QG. The first one was constructed by Simpson and Visser in Ref.~\cite{cha26}, which is a solution of Einstein’s equation when considering the coupling of the gravitational interaction with a phantom scalar field with a nonlinear electrodynamics \cite{chb32+,chb32}. They remove the singularity by replacing $r$ in the components of  Schwarzschild metric with ${\sqrt {{r^2} + {a^2}} }$, yielding an intriguing line element for the RBH (S-V black hole) that can be expressed as follows:
\begin{equation}
\label{eq2}
{\text{d}}{s^2} =  - \left( {1 - \frac{{2m}}{{\sqrt {{r^2} + {a^2}} }}} \right){\text{d}}{t^2} + {\left( {1 - \frac{{2m}}{{\sqrt {{r^2} + {a^2}} }}} \right)^{ - 1}}{\text{d}}{r^2} + \left( {{r^2} + {a^2}} \right){\text{d}}{\Omega ^2},
\end{equation}
where the parameter $a = {a_0}{\ell_{\text{pl}}}{\left( {{m \mathord{\left/ {\vphantom {m {{m_{\text{pl}}}}}} \right. \kern-\nulldelimiterspace} {{m_{\text{pl}}}}}} \right)^{\frac{1}{3}}}$ with the Planck mass ${m_{\text{pl}}}$ and Planck length ${l_{\text{pl}}}$ guarantees the regularization of the central region and reflects the effects of QG. ${\text{d}}{\Omega ^2} = {\text{d}}{\theta ^2} + {\sin ^2}\theta {\text{d}}{\phi ^2}$ is the metric of the unit sphere. In addition, $a_0$ is a dimensionless constant and  $a$ plays the role of controlling the geometric properties of spacetime. For $a < 2m$, it is a regular black hole which has two horizons at  ${r_ \pm } =  \pm \sqrt {{{\left( {2m} \right)}^2} - {a^2}} $. For $a = 2m$, the line element becomes a one-way wormhole with a null throat with an extremal event horizon. When $a > 2m$, one has a symmetric traversable wormhole with a two-way timelike throat of the Morris-Thorne type.

The second one was proposed in Refs.~\cite{cha27,chb27}. They incorporated a quantum gravity term into the interior of the horizon in Schwarzschild black hole, resulting in a singularity-free line element. This new solution can be expressed as follows:
\begin{equation}
\label{eq1}
{\text{d}}{s^2} =  - \frac{{4{{\left( {{\tau ^2} + l} \right)}^2}}}{{2m - {\tau ^2}}}{\text{d}}{\tau ^2} + \frac{{2m - {\tau ^2}}}{{{\tau ^2} + l}}{\text{d}}{x^2} + {\left( {{\tau ^2} + l} \right)^2}{\text{d}}{\Omega ^2},
\end{equation}
where $\tau$ and $x$ represent the temporal and spatial  coordinates within the horizon of black hole, respectively. The parameter $l \sim {a_0} {\ell_{\text{pl}}}{\left( {{m \mathord{\left/ {\vphantom {m {{m_{\text{pl}}}}}} \right. \kern-\nulldelimiterspace} {{m_{\text{pl}}}}}} \right)^{\frac{1}{3}}}$ is a characteristic length scale introduced by QG effects. The event horizon of the black hole is located at $\tau =\pm \sqrt {2m } $. It is evident that the presence of QG eliminates the singularity, allowing matter to travel through the central region of the black hole  at $\tau = 0$ and reach the white hole on the other side with  $\tau < 0$. Consequently, the information loss paradox due to Hawking radiation may potentially be resolved as information can cross the internal region without singularity.

Although the above RBHs succeed in removing the singularity, and most notably allowing matter to travel through the center of the black hole and reach the white hole region in other universe, unfortunately they suffer from some severe unsatisfactory points. For the metric in~Eq.~(\ref{eq2}), one has to  pay the price of violating energy conditions  everywhere, as discussed in \cite{cha29}. For the metric in Eq.~(\ref{eq1}), one has to face the discontinuity of the spacetime at the horizon whenever $l$ is a non-zero constant, if one assumes that  outside the horizon it is described by Schwarzschild metric with $l=0$, as performed in Refs.~\cite{cha27,chb27}. Or if one intends to extend (\ref{eq1}) to the region outside the horizon by inverse  coordinate transformation, then it would become

\vspace{-1cm}
\begin{equation}
 \label{eq1-1}
{\text{d}}{s^{\text{2}}} =  - \frac{{r - 2m}}{{r + l}}{\text{d}}{t^{\text{2}}} + \frac{{{{\left( {r + l} \right)}^2}}}{{r\left( {r - 2m} \right)}}{\text{d}}{r^{\text{2}}} + {\left( {r + l} \right)^2}{\text{d}}{\Omega ^{\text{2}}},
\end{equation}
which would violate energy conditions even at infinity, similarly as the metric in (\ref{eq2}). It is well known that energy conditions play a crucial role in the construction of black hole geometries. It is reasonable to expect energy conditions to be violated  near the center of the black hole due to the presence of QG effects. While, for the region far away from the center where effects of QG are weaken, one expects that ordinary energy conditions should be satisfied.  In order to address this issue, in this paper we intend to construct new black-to-white hole solutions through a surgical procedure. These solutions violate energy conditions only in the vicinity of the central region, while the continuity of the metric can be guaranteed up to the arbitrary order which is controlled by the power factor $n$.

The work is organized as follows. In the next section, we discuss the general conditions leading to the violation of energy conditions under the static spacetime with spherical symmetry. Then we propose a framework for constructing black-to-white holes with sub-Planckian curvature, where the energy conditions are violated only within a small region of space time. The geodesics of particles will also be discussed and the completeness of the geodesics is confirmed in the Carter-Penrose diagram with the maximal extension of the spacetime. Then, the metric of black-to-white hole in the interior of horizon presented in section~\ref{sec3} is modified such that it is continuously connected with the Schwarzschild metric outside the horizon. Our results and conclusions are presented in section~\ref{sec4}. To simplify the notation,  in this work we adopt the Planck units $c=k_B=G=\hbar=1$ and the convention with signature $(-,+,+,+)$.

\section{New black-to-white hole solutions I}
\label{sec2}
\subsection{Energy conditions in general spherically symmetric RBH}
\label{sec2-1}
We consider a general static and spherically symmetric line element with effects of QG which is given by
\begin{align}
\label{eq3}
\text{d}{s^2} =  - A\left( {r,a} \right){\text{d}}{t^2} + B{\left( {r,a} \right)^{ - 1}}{\text{d}}{r^2} + {g^2}\left( {r,a} \right){\text{d}}{\Omega ^2},
\end{align}
where $A\left( {r,a} \right)$, $B\left( {r,a} \right)$ and $g\left( {r,a} \right)$ are three general functions with the parameter $a$ representing the strength of QG. 
For simplicity, we set $A\left( {r,a} \right)=B\left( {r,a} \right)=f\left( {r,a} \right)$. Thus, the event horizon is specified by the equation $f\left( {r,a} \right)=0$. Following Ref.~\cite{cha6}, a regular black hole  can be justified by checking its curvature invariants. Thus we derive the following three curvature invariants which are most frequently used, namely the Ricci scalar, the contraction of Ricci tensor and Kretschmann scalar:
\begin{equation}
\label{eq4}
R = \frac{2}{{{g^2}}}\left( {1 - f{{g'}^2}} \right) - \frac{4}{g}\left( {f'g' - fg''} \right) - f'',
\end{equation}
\begin{equation}
\label{eq5}
\begin{gathered}
  {R_{\mu \nu }}{R^{\mu \nu }} = \frac{1}{2}{{f''}^2} + \frac{2}{{{g^2}}}\left( {2{{f'}^2}{{g'}^2} + 3{f^2}{{g''}^2} + 4ff'g'g''} \right) + \frac{2}{g}\left( {f'g'f'' + ff''g''} \right) \\
   + \frac{4}{{{g^3}}}\left( {{f^2}{{g'}^2}g'' + ff'{{g'}^3} - fg'' - f'g'} \right) + \frac{2}{{{g^4}}}\left( {{f^2}{{g'}^4} - 2f{{g'}^2} + 1} \right), \\
\end{gathered}
\end{equation}
\begin{equation}
\label{eq6}
{K^2}= R_{\mu\nu\gamma \varphi}R^{\mu\nu\gamma \varphi} = \frac{4}{{{g^2}}}\left( {{{f'}^2}{{g'}^2} + 2{f^2}{{g''}^2} + 2ff'g'g''} \right) + \frac{4}{{{g^4}}}\left( {{f^2}{{g'}^4} - 2f{{g'}^2} + 1} \right) + {f''^2},
\end{equation}
where the prime represents the derivative with respect to the coordinate $r$. From Eqs.~(\ref{eq4})-(\ref{eq6}), one finds that to generate a regular spacetime, there are three schemes that make the curvature invariant not divergent at the center of the black hole: (1) Taking $g=r$ and ensuring that the  term with parameter $a$ in  $f$ affects the denominator such that the ratio is not divergent at $r = 0$.  Bardeen and Hayward black holes are such prominent examples and more extensions with exponential Newtonian potential are given in Refs.~\cite{chz1,chz2,chz3,chz4}; (2) $f$ could be a general function with horizon  and one requires that $ g\left( {r,a} \right) \ne 0$ at the center of the black hole, and the first and second order derivatives of all free functions with coordinate $r$ are finite everywhere~\cite{cha29}. (3) One may simultaneously add correction terms into $f$ and $g$ to ensure that they are not equal to zero at $r = 0$, such as Refs.~\cite{chx29,chx30}. Obviously, the S-V black holes belong to the second category that the non-zero parameter $a$ prevents the denominator of the curvature invariants from divergent at $r = 0$, and thereby avoids singularity.

Next, let us discuss null energy condition (NEC). Based on the line element~(\ref{eq3}) and  Einstein's equation $G_\nu ^\mu  = 8\pi GT_\nu ^\mu$, the non-zero components of stress-energy tensor are given by
\begin{equation}
\label{eq7-1}
T_t^t = \frac{1}{{8\pi }}\left[ {\frac{1}{g}\left( {f'g' + 2fg''} \right) - \frac{1}{{{g^2}}}\left( {1 - f{{g'}^2}} \right)} \right],
\end{equation}
\begin{equation}
\label{eq7-2}
T_r^r  = \frac{1}{{8\pi }}\left( {\frac{{f'g'}}{g} - \frac{{1 - f{{g'}^2}}}{{{g^2}}}} \right),
\end{equation}
\begin{equation}
\label{eq7-3}
T_\theta ^\theta  = T_\phi ^\phi  = \frac{1}{{8\pi }}\left[ {\frac{1}{g}\left( {f'g' + fg''} \right) + \frac{1}{2}f''} \right].
\end{equation}
For spacetime outside the horizon, the energy density $\rho$, pressure $p_i$ can be read from the stress-energy tensor $T_\nu ^\mu$ as $T_\nu ^\mu  = {\text{diag}}\left[ { - \rho ,{p_{//}},{p_ \bot },{p_ \bot }} \right]$, while inside the horizon, the relation becomes $T_\nu ^\mu  = {\text{diag}}\left[ { - {p_{//}},\rho ,{p_ \bot },{p_ \bot }} \right]$~\cite{cha29}. To make the NEC hold, it is required that $\rho  + {p_{//}} \geqslant 0$ and $\rho  + {p_ \bot } \geqslant 0$, leading to
\begin{equation}
\label{eq8}
\rho  + {p_{//}} =  \pm \frac{{fg''}}{{4\pi g}}, \quad \rho  + {p_ \bot } = \frac{{2 - 2f{{g'}^2} + {g^2}f'' \pm 2fgg''}}{{16\pi {g^2}}},
\end{equation}
where the negative sign corresponds to the case $f\left( {r,a} \right) > 0$, while the positive sign corresponds to the case $f\left( {r,a} \right) < 0$. For spacetime with ordinary angular components of the metric, namely $g=r$ and $g''=0$, then $\rho + {p_{//}} =0$, $\rho  + {p_ \bot } = (2 - 2f + {r^2}f'')/(16\pi {r^2})$, whose behavior depends on the specific form of $f\left(r,a\right)$. If one intends to construct a RBH by considering a general form of angular components of the metric as $g\left( {r,a} \right)$, then the above analysis shows that if $g\left( {r,a} \right)>0$ and ${g''\left( {r,a} \right)}>0$, Eq.~(\ref{eq8}) is always negative, regardless of whether it is inside or outside the horizon~\cite{cha29}, and the function $g\left( {r,a} \right)$ of the S-V black hole has exactly this property since $g=\sqrt{r^2+a^2}$ with $a\ne 0$, giving rise to the violation of NEC everywhere in the S-V spacetime. To overcome this difficulty, based on the above analysis we propose to modify the function  $g\left( {r,a} \right)$ to make the null energy condition violated only close to the center of the black hole, and this will be what we intend to do in next subsections.
\vspace{-0.6cm}

\subsection{The metric form of the new black-to-white solution}
\label{sec2-2}
We propose a scheme to construct the black-to-white holes by modifying the function $g\left( {r,a} \right)$. The goal of this work is to ensure that the resulting black hole satisfies the following requirements. (i) the spacetime is regular everywhere due to  effects of QG, thereby ensuring that the curvature variable remains finite and sub-Planckian; (ii) the violations of energy conditions are limited to the vicinity of the core of holes, while in the region far from the center of holes all the energy conditions are satisfied; (iii) striving for the integrity of the geodesic line, ensuring a smooth and unobstructed flow of material within the black-to-white holes; (iv) enabling matter to pass from a black hole in our universe into a white hole in another universe, thereby upholding the crucial principle of information conservation within the black hole. In fact, obtaining a black-to-white solution that meets the four requirements above is highly challenging. For instance, in Ref.~\cite{chb32}, it was proven that for the S-V black hole, when $a$ is non-zero in the function $g\left( {r,a} \right)$, the NEC is violated throughout. It means that there is no room for one to construct new smooth solutions with improved energy conditions with usual schemes. One potential approach to overcome this limitation is the gluing scheme, which is utilized in general relativity literature to create spacetimes with rich structure~\cite{chx31,chx32,chx33,chx34,chx35,chx36}. Inspired by this approach, we can decompose a black-to-white hole solution into two segments intersecting at a gluing surface. The ``internal" segment approaching $r = 0$ is influenced by effects of QG, and its spacetime can be described by a known line element, such as Eq.~(\ref{eq2}) or Eq.~(\ref{eq1}), resulting in energy conditions violation, elimination of singularity, and enabling matter traversal towards the white hole. Conversely, in the ``outer" segment away from $r = 0$, where effects of QG diminish, the spacetime gradually transitions to the standard Schwarzschild case. Additionally, ensuring the continuity of the resultant spacetime and its compliance with Einstein's equation requires consideration of joint conditions at the gluing surface. Since one side of the gluing surface corresponds to the a known metric, the form of the metric on the other side can be derived based on the order of the continuity that one wants. Under these stringent criteria, the black-to-white solution can be viewed as extension of the line element in Eq.~(\ref{eq2}) or Eq.~(\ref{eq1}), thereby constituting solution to Einstein's equation. For the purposes above, we propose to introduce a gluing surface inside the horizon of space time and let the function $g\left( {r,a} \right)$ in Eq.~(\ref{eq3}) take different forms on both sides of the gluing surfaces. Specifically, we propose it takes the following form

\vspace{-\baselineskip}
\begin{subequations}
\label{eq9}
\begin{numcases}{{g^2}\left( {r,a} \right) =}
{r^2} + {a^2}{\left( {1 - \frac{{{r^2}}}{{{\varepsilon ^2}}}} \right)^n}, & for ${r   \leq \varepsilon }$,\\
{r^2}, & for ${r > \varepsilon}$,
\end{numcases}
\end{subequations}
where $\varepsilon$ denotes the position of the gluing surface and $n$ is the power of the correction term due to effects of QG. Firstly, we remark that the position of gluing surface  $\varepsilon$ signals the place that the QG effects start to play a role, therefore one may expect it would have the same order as the parameter $a$, but mathematically it could be a different and independent parameter. Obviously, Eq.~(\ref{eq9}) indicates that the effect of QG plays a role only within the gluing surface (i.e. ${r \leq \varepsilon}$). Specially, at $r=0$ it becomes the form of $g=\sqrt {r^2+a^2}$ as in S-V black hole. While for ${r > \varepsilon}$ the effect of QG is so small that can be neglected and thus ${g^2}\left( {r,a} \right)$ reverts to the standard  Schwarzschild case, namely $g=r$. In the more general scenario, $A(r,a)$ may not be identical to $B(r,a)$ for $r > \varepsilon$. Their formulations are contingent upon the external spacetime intended for amalgamation in our investigation. Subsequently, we will formulate a new black-to-white hole solution rooted in the V-S spacetime, characterized by $A=B$ \cite{cha26}. Conversely, in section~\ref{sec3}, the novel solution derived from Refs.~\cite {chb27,cha27} engenders $A \ne B$. Secondly, it is easy to check that the continuity of the metric and its derivatives with respect to $r$  at gluing surface can be guaranteed up to the arbitrary order which is controlled by the power factor $n$. The larger $n$ is, the smoothness of the metric at gluing surface becomes better. Throughout this paper we specify $n =3$ such that the curvature invariants as well as the energy density and pressure are continuous at the gluing surface, since these quantities involve in the second derivative of the metric. Thirdly, it is interesting to notice that the correction term in Eq.~(\ref{eq9}) contains the even powers of $r$ only such that the modified metric remains to be symmetric for $r\rightarrow -r$. This property also plays an essential role in the construction of black-to-white hole solutions.

Now, by considering $ f\left( {r,a} \right)=1 - {{2m} \mathord{\left/ {\vphantom {{2m} {\sqrt {{r^2} + {a^2}} }}} \right. \kern-\nulldelimiterspace} {\sqrt {{r^2} + {a^2}} }}$ as given in Eq.~(\ref{eq2}), we construct a new RBH whose metric takes the following form
\begin{equation}
\label{eq10}
{{\text{d}}{s^2} =  - \left( {1 - \frac{{2m}}{{\sqrt {{r^2} + {a^2}} }}} \right){\text{d}}{t^2} + {{\left( {1 - \frac{{2m}}{{\sqrt {{r^2} + {a^2}} }}} \right)}^{ - 1}}{\text{d}}{r^2} + g^2\left( {r,a} \right){\text{d}}{\Omega ^2}},
\end{equation}
where $a = {a_0}{\ell_{{\text{pl}}}}{\left( {{m \mathord{\left/ {\vphantom {m {{m_{\text{pl}}}}}} \right.  \kern-\nulldelimiterspace} {{m_{\text{pl}}}}}} \right)^{\frac{1}{3}}}$ is the QG parameter with the adjustable dimensionless coefficient $a_0$.
The range  of the coordinates are $t \in \left( { - \infty ,\infty } \right)$, $\theta  \in \left. {\left( {0,\pi } \right.} \right]$, $\phi  \in \left. {\left( { - \pi ,\pi } \right.} \right]$, and  by considering the geometric completeness along radial direction, we  also extend the radial coordinate to $r \in \left( { - \infty ,\infty } \right)$. First of all, it is noticed that only even powers of $r$ appears in the metric, thus the geometry of space time is symmetric under the reflection of $r\rightarrow -r$.

\begin{figure}[htbp]
\setlength{\abovecaptionskip}{-0.2cm}
\centering
\includegraphics[width=0.6\textwidth]{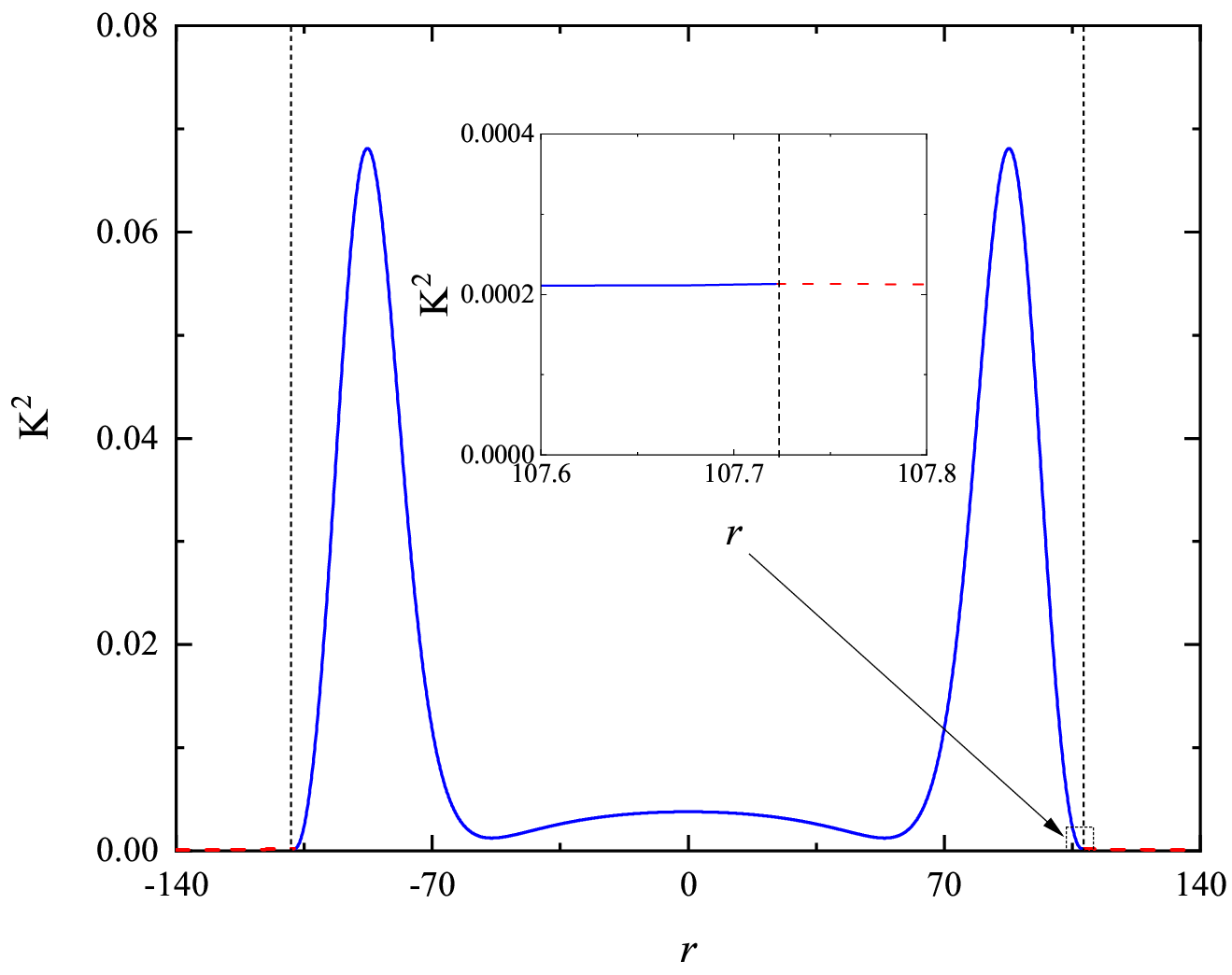}
\caption{The Kretschmann scalar $K^2$ as a function of  radial coordinate $r$ for $a = 10 {m^{\frac{1}{3}}}$ and $m = 10^4$ (in Planck units).}
\label{fig1}
\end{figure}

Next we are concerned with the curvature invariants of this spacetime. Substituting the components of Eq.~(\ref{eq10}) into Eq.~(\ref{eq6}), one obtains the Kretschmann scalar curvature, which is presented in Appendix~\ref{appA}. In particular, at the center of the black hole, the Kretschmann scalar is
\begin{equation}
\label{eq11+}
{K^2}\left( {r = 0} \right) = \frac{{36{m^2}}}{{{a^6}}} + \frac{{288{m^2}}}{{{a^2}{\varepsilon ^4}}} + \frac{{12}}{{{a^4}}} - \frac{{32m}}{{{a^5}}} + \frac{{72}}{{{\varepsilon ^4}}} - \frac{{288m}}{{a{\varepsilon ^4}}} - \frac{{48}}{{{a^2}{\varepsilon ^2}}} + \frac{{192m}}{{{a^3}{\varepsilon ^2}}} - \frac{{192{m^2}}}{{{a^4}{\varepsilon ^2}}}.
\end{equation}
First of all, due to the presence of parameter $a$ and $\varepsilon$, we notice that  the Kretschmann scalar curvature can be finite rather than divergent. Specially, if one sets $a\sim \varepsilon \sim m^{1/3}$, the leading terms may be mass independent and the subleading terms are inversely proportional to the mass of black holes, which provides a scheme to maintain the curvature to be bound by the energy density at Planck scale. Notably, in the Planck units, one has ${\ell_{{\text{pl}}}} = {m_{{\text{pl}}}} = 1$, which implies that the parameter $a$ appearing in the metric~(\ref{eq2}) have the dimension of length, while $m$ has the dimension of mass or energy. It means that the parameter $a$ should be understood as
 $a = {a_0}{\ell_{{\text{pl}}}}{\left( {{m \mathord{\left/
 {\vphantom {m {{m_{{\text{pl}}}}}}} \right.
 \kern-\nulldelimiterspace} {{m_{{\text{pl}}}}}}} \right)^{\frac{1}{3}}}$, where $a_0$ is a dimensionless coefficient. At the same time $m$ should be understood as a quantity with the dimension of mass, with $m_{{\text{pl}}}$ as the unit. Thus, it is sufficient that  $a_0$ is taken such that the curvature and energy are below $1$. In the study later on, without loss of generality, we set $a = a_0 {m^{\frac{1}{3}}}$ and $\varepsilon =a/2$ with $a_0=10$ and the mass independence of the scalar curvature $m=10^4$. In these settings, the  Kretschmann scalar curvature as the function of $r$ is depicted in Fig.~\ref{fig1}

In this figure, the position of gluing surface is indicated by vertical dashed lines at $r=\varepsilon \approx \pm 107.72$. The value of the Kretschmann scalar curvature is plotted as the red dashed curve based on Eq.~(\ref{eqa1-2})  which is applicable for the region outside the gluing surface. In this region, the effects of QG can be neglected, and thus the red dashed curve asymptotically tends to zero as $r$ goes to infinity. Within the region inside the gluing surface, the behavior of $K^2$ is described by the blue solid curve based on Eq.~(\ref{eqa1-1}). Firstly, we remark that  the scalar curvature exhibits the continuous behavior at the gluing surface indeed, where $K^2 \approx 0.0002$. Secondly, we point out that the scalar curvature  $K^2$ now is finite at $r=0$, implying the absence of the singularity at the center of the black hole. Furthermore, we remark that the maximal value of the scalar curvature $K^2$ locates the position near the gluing surface, which is in contrast to case in the original RBH where the maximal value of $K^2$ always appears at the center of the black hole, namely $r=0$.

\begin{figure}[htbp]
\centering
\includegraphics[width=0.55\textwidth]{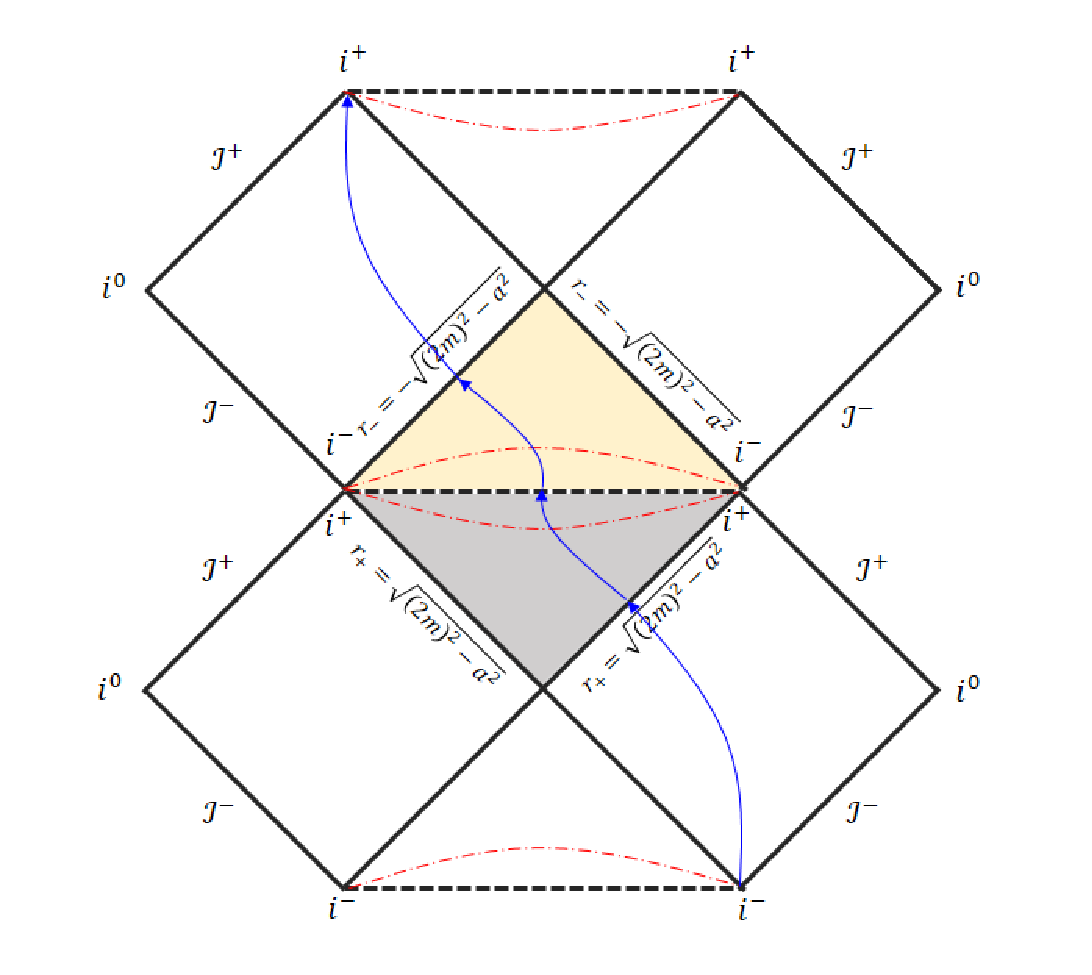}
\caption{Carter-Penrose diagram for the maximally extended  spacetime for $a \in \left( {0,2m} \right)$.}
\label{fig2}
\end{figure}

Next in order to gain a more comprehensive understanding on the absence of the singularity in this spacetime,  we turn to the completeness of geodesics, particularly near the region inside the gluing surface. Thanks to the spherical symmetry of the spacetime, here we restrict our attention to particles with the motion along the radial direction. Following the viewpoint in Refs.~\cite{cha30,chb30}, the radial geodesic equation of motion is given by
\begin{equation}
\label{eq12+}
{g_{tt}}{\left( {\frac{{{\text{d}}t}}{{{\text{d}} \chi}}} \right)^2} + {g_{rr}}{\left( {\frac{{{\text{d}}r}}{{{\text{d}} \chi}}} \right)^2} =  - {\delta },
\end{equation}
where $\chi$ is the affine parameter, and $\delta =0$ for massless particles, while $\delta=1$ for massive particles.
By virtue of the expression, $E =  - {g_{tt}}\left( {{{{\text{d}}t}\mathord{\left/ {\vphantom {{{\text{d}}t} {{\text{d}}\lambda }}} \right. \kern-\nulldelimiterspace} {{\text{d}}\chi}}} \right)$, the variation of the affine parameter along the geodesics can be integrated as
\begin{equation}
\label{eq13+}
\Delta \chi  = \int_{{r_p}}^{r_q} {\sqrt { - \frac{{{g_{tt}}{g_{rr}}}}{{\delta {g_{tt}} + {E^2}}}} {\text{d}}r},
\end{equation}
where ${r_p}$ and ${r_q}$ denote the final  and initial positions of motion, respectively. Substituting Eq.~(\ref{eq10}) into Eq.~(\ref{eq13+}), the affine parameter for massless particles can be expressed as follows
\begin{equation}
\label{eq14+}
\Delta \chi  = \int_{{r_p}}^{{r_q}} {\frac{\text{d}r}{E}} = \frac{1}{E }\left( {{r_q} - {r_p}} \right).
\end{equation}
It is clear that the affine parameter is only related to the coordinate $r$ and the energy $E$. On the other hand, the affine parameter  for massive particles is
\begin{equation}
\label{eq15+}
\Delta \chi  = \int_{{r_p}}^{{r_q}} {{{\left[ {\frac{{\sqrt {{a^2} + {r^2}} }}{{2m + \sqrt {{a^2} + {r^2}} \left( {{E ^2} - 1} \right)}}} \right]}^{\frac{1}{2}}}{\text{d}}r},
\end{equation}
In Eq.~(\ref{eq14+}) and Eq.~(\ref{eq15+}), since $m$, $E$ and the integration interval are finite, the affine parameter $\Delta \chi$ remains finite when reaching the center of the black hole. This indicates that the spacetime is incomplete at $r = 0$, and it is possible to extend the spacetime beyond $r = 0$ to negative values. In fact, a simple analysis reveals that $\Delta \chi$ diverges only when $r$ approaches infinity, demonstrating that the spacetime is complete in the region $r \in (-\infty, +\infty)$. Based on the results obtained above, we can  plot the Carter-Penrose diagram for the maximally extended  spacetime, as illustrated in Fig.~\ref{fig2}. The red dashed line denotes the location of the gluing surface, where the impact of QG starts to become significant. Importantly, at $r = 0$, QG creates a pathway through which matter can travel (blue line) from the region of the black hole (grey diamond) to the white hole (yellow diamond) region.

\subsection{ The stress-energy tensor and energy conditions}
\label{sec2-3}
In this subsection we focus on the energy conditions in this RBH. Based on Eqs.~(\ref{eq7-1})-(\ref{eq8}), we obtain the non-zero components of stress-energy tensors inside and outside the gluing surfaces, which are presented in Eqs.~(\ref{eqb1-1})-(\ref{eqb1-4})  of Appendix~\ref{appB}. In the region $r_{H} < r < \infty$, the energy density $\rho$ and the pressures $p_i$ are given by $\rho = -T_t^t$, $p_{//} = T_r^r$, and $p_{\perp} = T_\theta^\theta = T_\phi^\phi$. However, in the region $\varepsilon < r < r_{H}$, the roles of the $t$ and $r$ coordinates are swapped, leading to $\rho = -T_r^r$, $p_{//} = T_t^t$, and $p_{\perp} = T_\theta^\theta = T_\phi^\phi$. For the sake of convenience, here we consider a simple special case of  $\varepsilon={a \mathord{\left/ {\vphantom {a 2}} \right. \kern-\nulldelimiterspace} 2}$, and results are presented in Eqs.~(\ref{eqb1-4})-(\ref{eqb1-7})  of Appendix~\ref{appB}.

\begin{figure}[htbp]
\setlength{\abovecaptionskip}{-0.5cm}
\setlength{\belowcaptionskip}{-0.3cm}
\centering
\subfigure{
\begin{minipage}{0.42\textwidth}
\includegraphics[width=\textwidth]{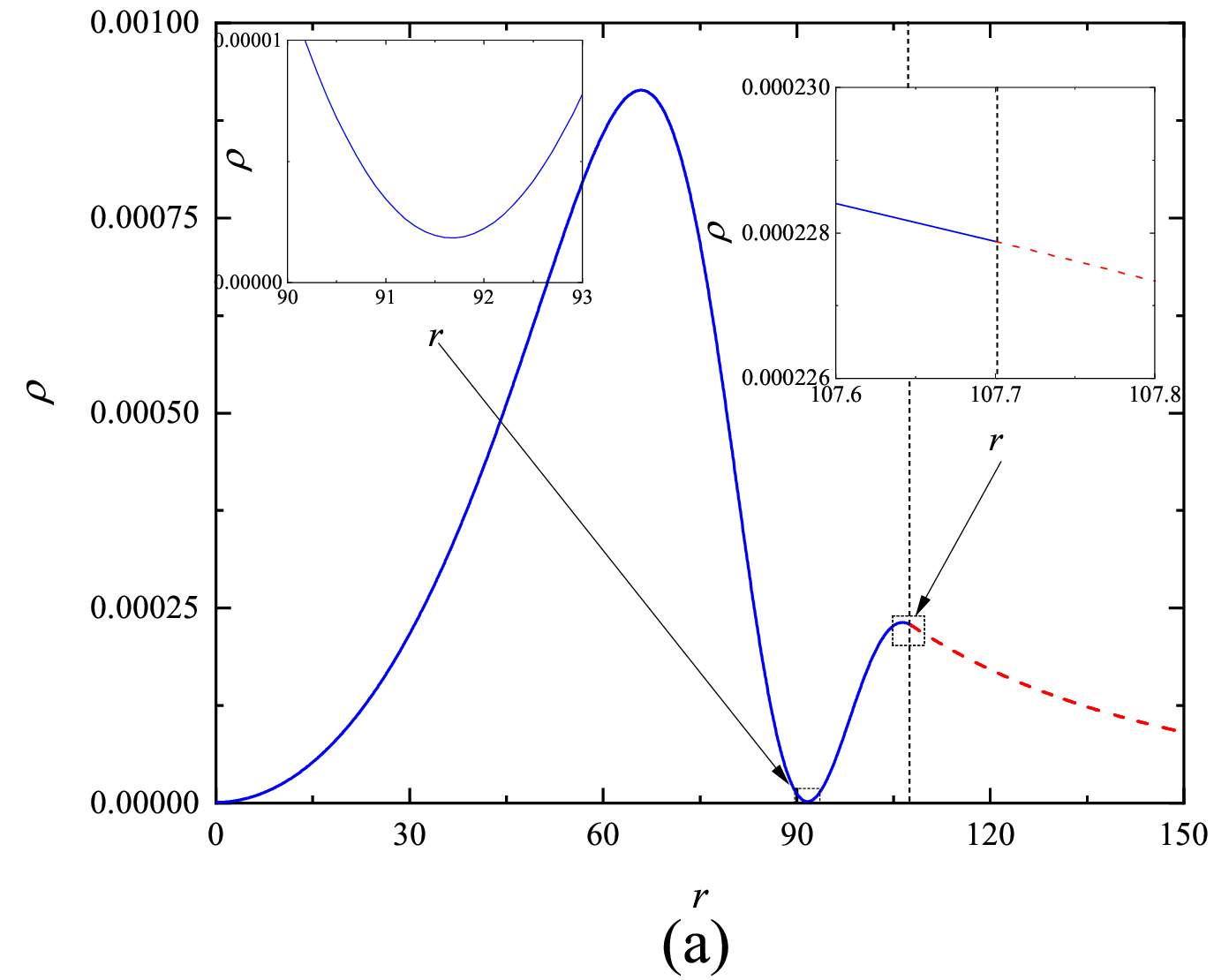}
\label{fig3-a}
\end{minipage}
}
\subfigure{
\begin{minipage}{0.42\textwidth}
\includegraphics[width=\textwidth]{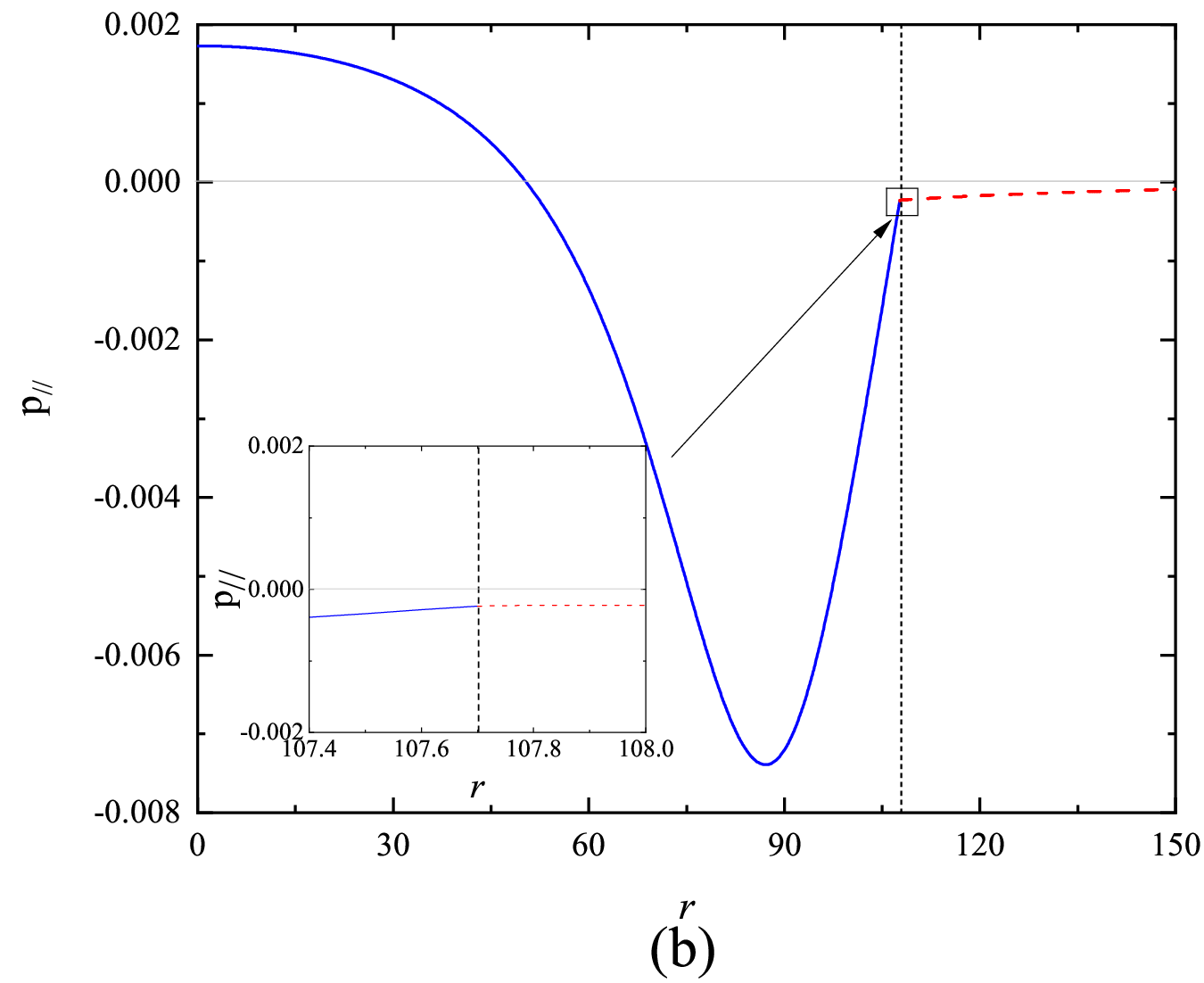}
\label{fig3-b}
\end{minipage}
}
\subfigure{
\begin{minipage}{0.42\textwidth}
\vspace{-0.7cm}
\includegraphics[width=\textwidth]{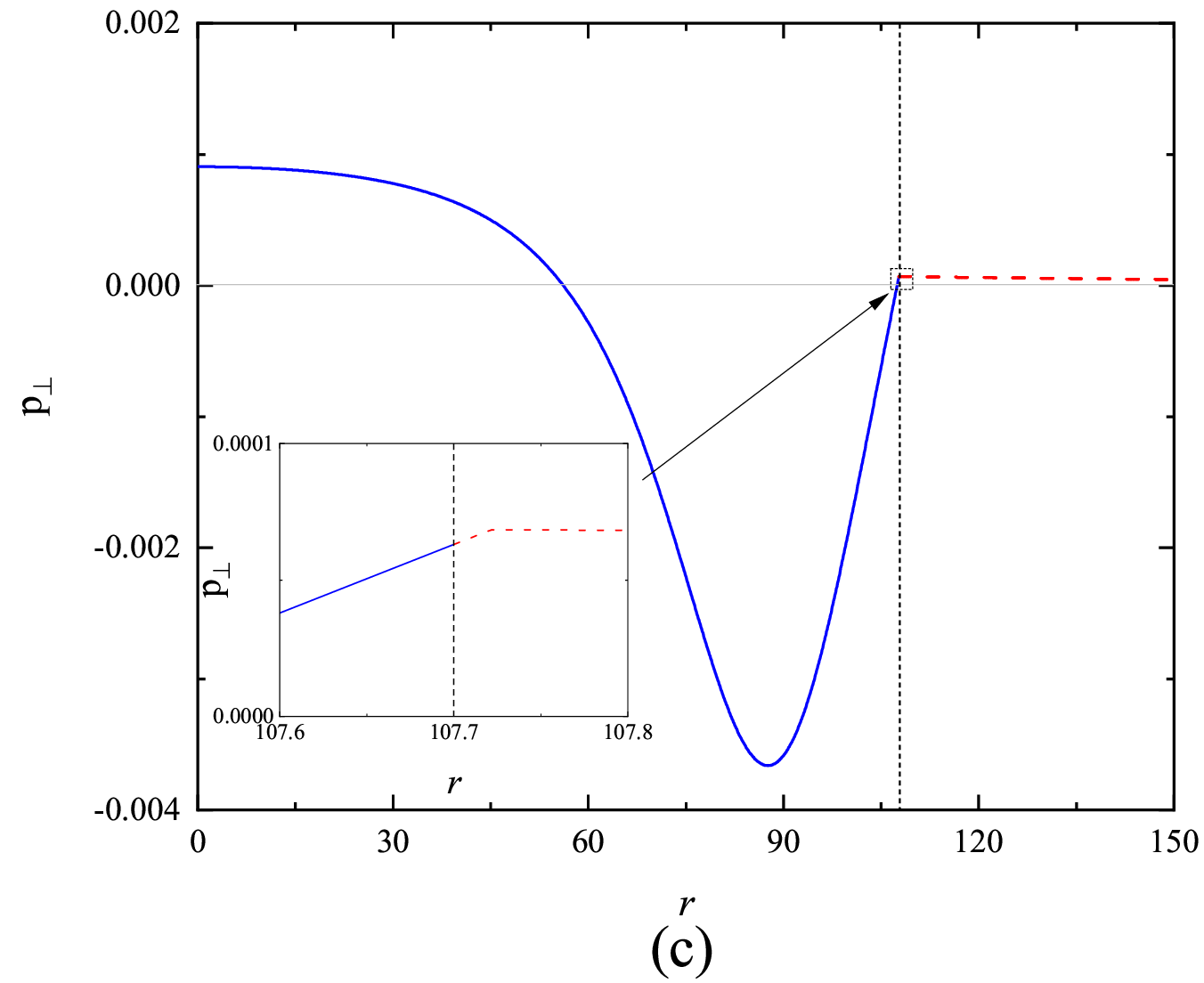}
\label{fig3-c}
\end{minipage}
}
\caption{The behavior of the energy density $\rho$ and pressure $p_{i}$. (a) $\rho  $ versus $r$. (b) $ {p_ {//} }$ versus $r$. (c) $ {p_ \bot }$ versus $r$. Here $a = 10 {m^{\frac{1}{3}}}$ and $m = 10^4$ (in Planck units).}
\label{fig3}
\end{figure}

As we are discussing the region within the event horizon, the energy density $\rho$, pressure ${p_{//}}$ and  $p_{\perp}$ are given by $- T_r^r$, $T_t^t$, and $T_\theta^\theta = T_\phi^\phi$, respectively. Generally,  it is reasonable to treat the matter in spacetime of black holes as an ideal fluid, which implies that the energy density and pressure should be continuous throughout the spacetime. From the obtained expressions for $\rho$ and $p_i$, we illustrate the behavior of the energy density and the pressure in Fig.~\ref{fig3}. The values of quantities inside the gluing surface are plotted with blue solid curves, while the values of quantities outside the gluing surface are plotted with red dashed curves. The insets clearly demonstrate the continuity of these curves at the gluing surface, indicating the continuous feature of the spacetime geometry.

Next, we analyze the energy conditions over the spacetime. For convenience, we list the inequalities that the energy and the pressure should satisfy under various energy conditions in Tab.~\ref{tab1}, including the NEC,  weak energy condition (WEC), dominant energy condition (DEC) and strong energy conditions (SEC).
\begin{table}[htbp]
\centering
\caption {\label{tab1} The  energy conditions and their relationship with energy density and pressure.}
\begin{tabular}{cc}
\hline
Energy conditions & The relationship with density and energy   \\
\hline
NEC     &$\rho  + {p_{i}} \geq 0$\\
WEC   &$\rho \geq 0$,                       $\rho  + {p_{i}} \geq 0$ \\
DEC    & $\rho \geq 0$,                       $\rho  - \left| {{p_i}} \right| \geq 0$ (or $\rho  -  {{p_i}}  \geq 0$ and $\rho  +  {{p_i}}  \geq 0$) \\
SEC    & $\rho  + {p_{i}} \geq 0$,    $\rho  + {p_{//}} + 2 {p_{\bot}} \geq 0$ \\
\hline
\label{tab1}
\end{tabular}
\end{table}

First of all, let us discuss the NEC. In the region $r \in \left( {\varepsilon ,\infty } \right)$, the NEC is given by
\begin{equation}
\label{eq13}
\rho  + {p_{//}} = 0, \quad \rho  + {p_ \bot } = \frac{{{a^2}m\left( {2{a^2} + 5{r^2}} \right)}}{{8\pi {r^2}{{\left( {{a^2} + {r^2}} \right)}^{5/2}}}}.
\end{equation}
Obviously, the aforementioned formulations satisfy the requirements of the NEC. Subsequently, we examine the NEC within the region of gluing surface. Two simple algebras show that the NEC can be expressed as

\begin{figure}[H]
\setlength{\abovecaptionskip}{-0.5cm}
\setlength{\belowcaptionskip}{-0.3cm}
\centering
\subfigure{
\begin{minipage}{0.42\textwidth}
\includegraphics[width=\textwidth]{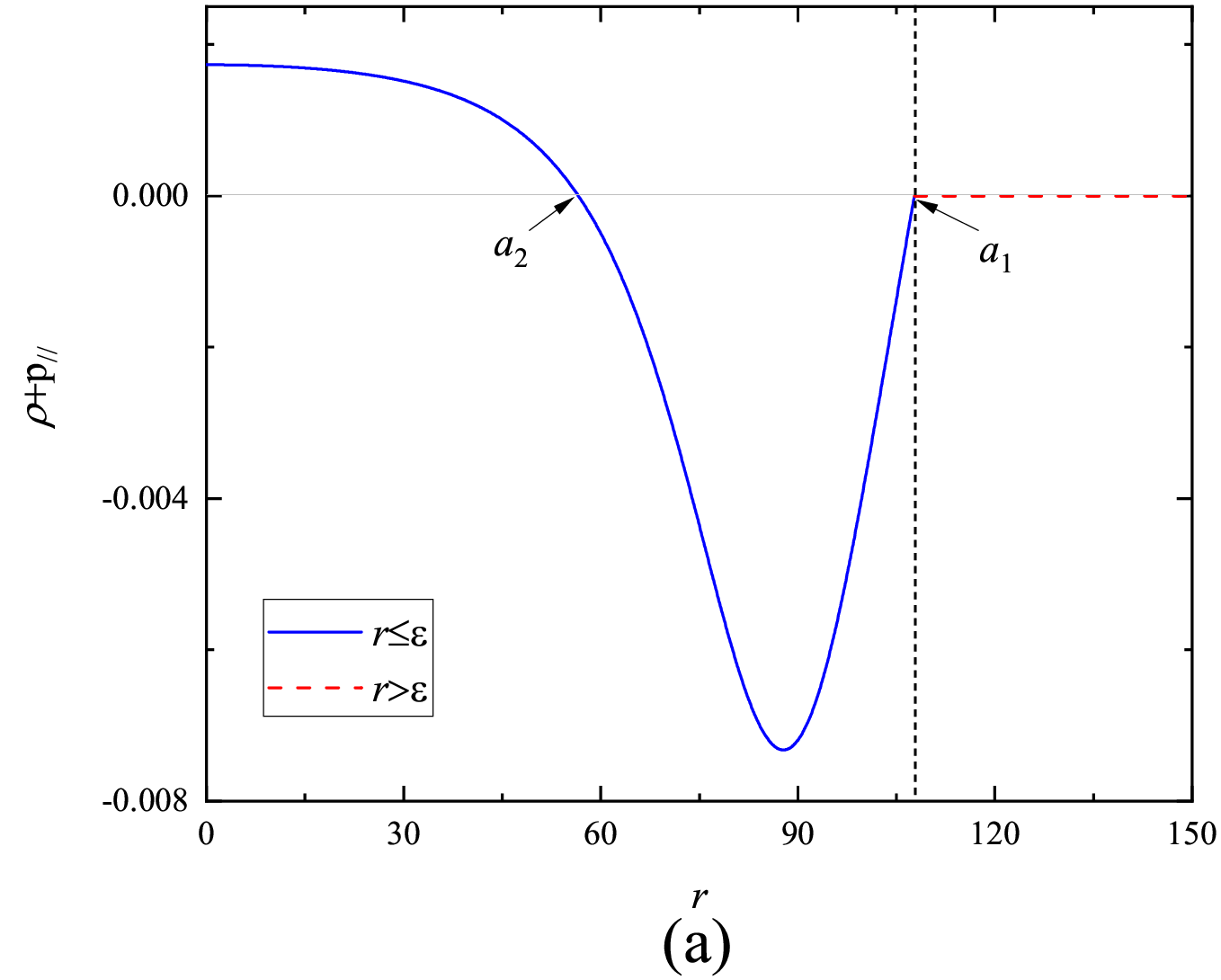}
\label{fig4-a}
\end{minipage}
}
\subfigure{
\begin{minipage}{0.42\textwidth}
\includegraphics[width=\textwidth]{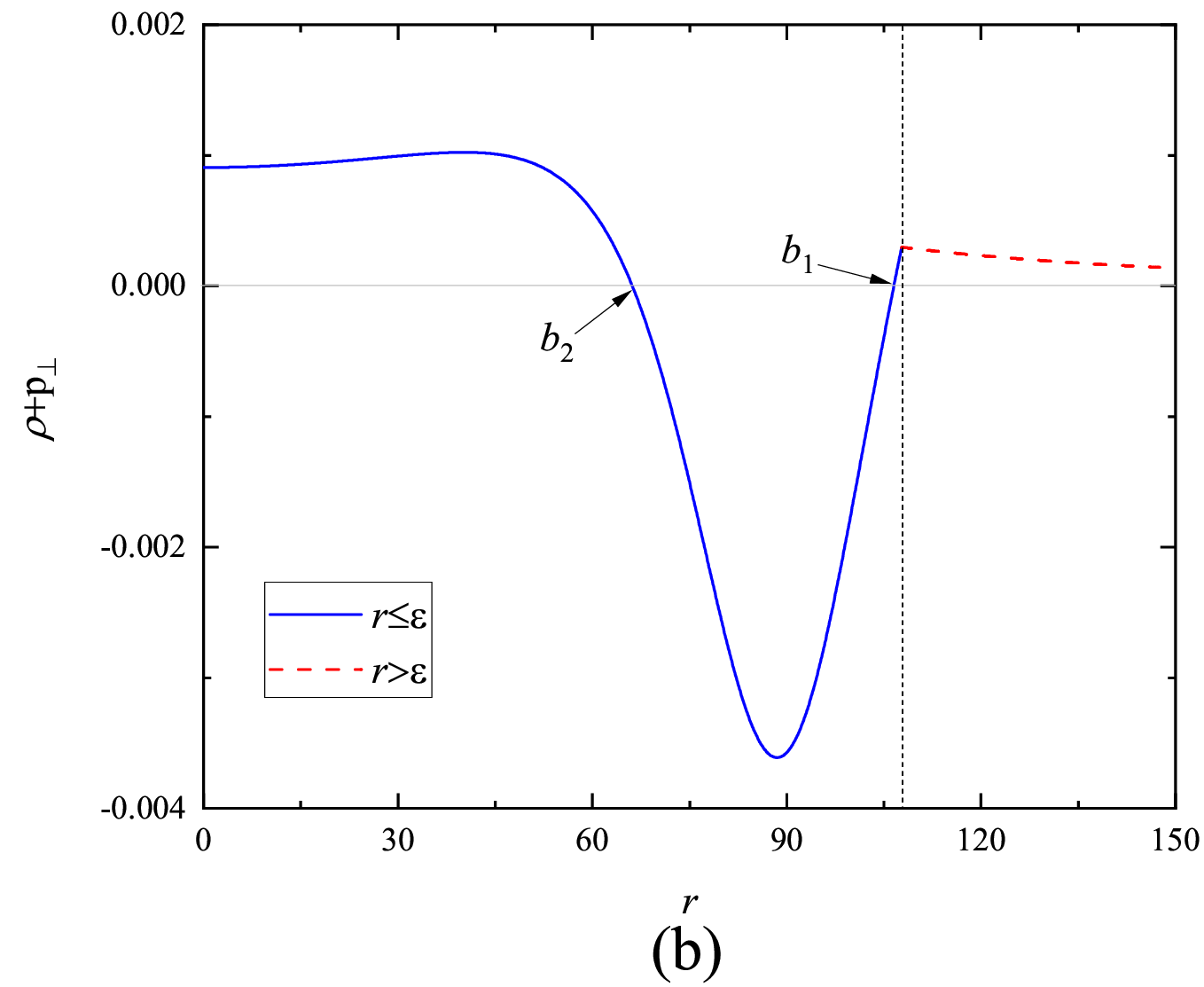}
\label{fig4-b}
\end{minipage}
}
\subfigure{
\begin{minipage}{0.42\textwidth}
\vspace{-0.7cm}
\includegraphics[width=\textwidth]{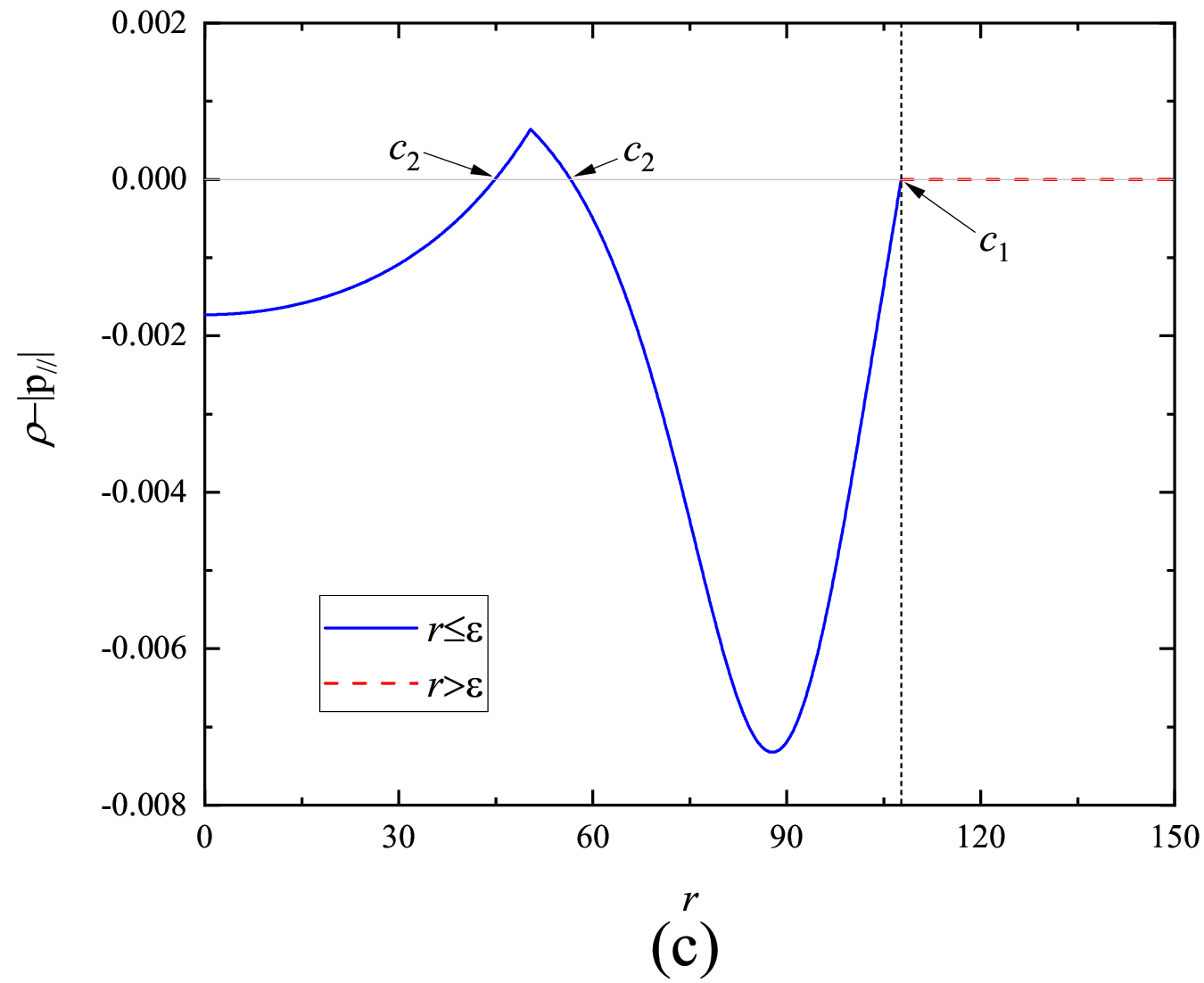}
\label{fig4-c}
\end{minipage}
}
\subfigure{
\begin{minipage}{0.42\textwidth}
\vspace{-0.7cm}
\includegraphics[width=\textwidth]{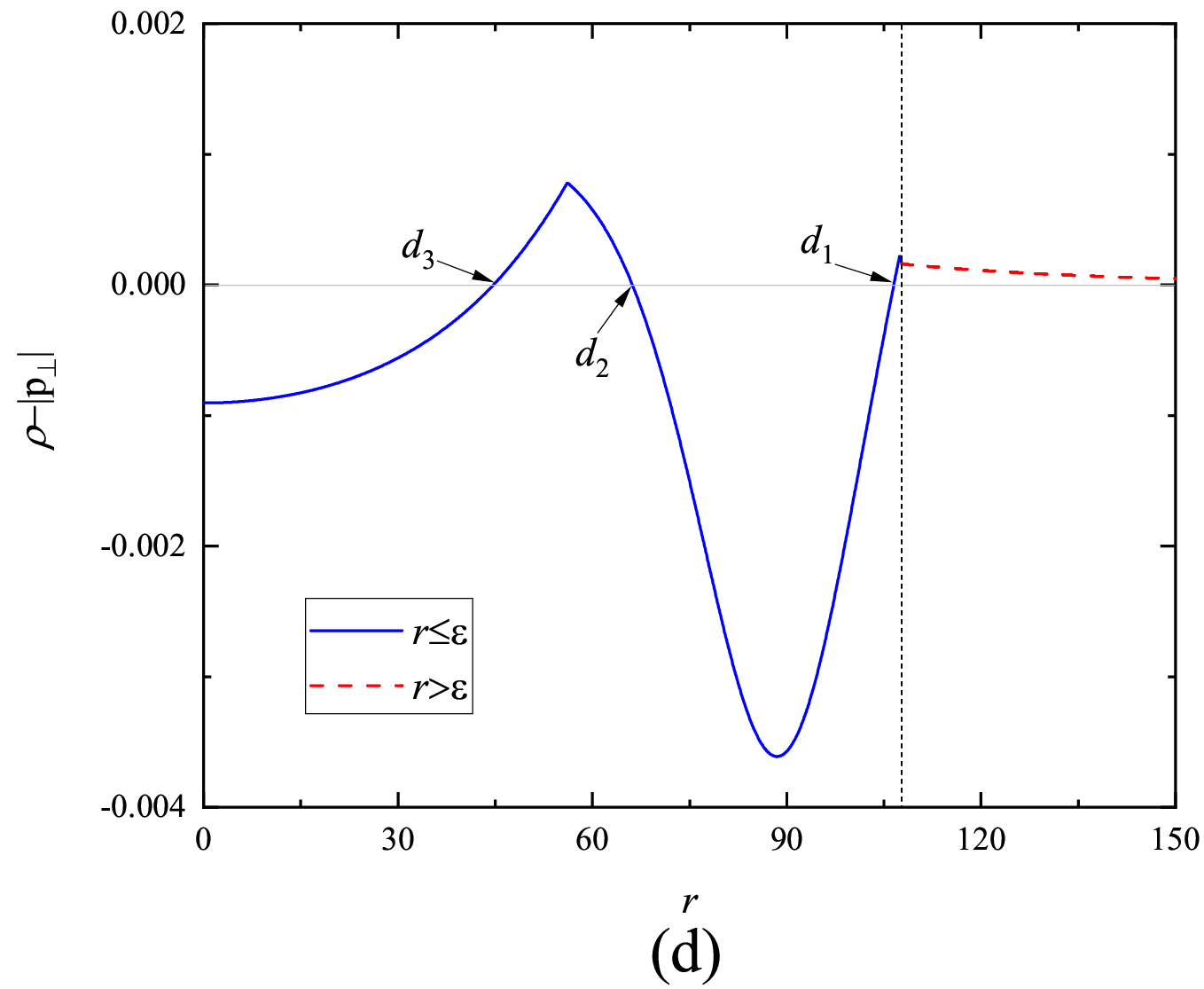}
\label{fig4-d}
\end{minipage}
}
\subfigure{
\begin{minipage}{0.421\textwidth}
\vspace{-0.7cm}
\includegraphics[width=\textwidth]{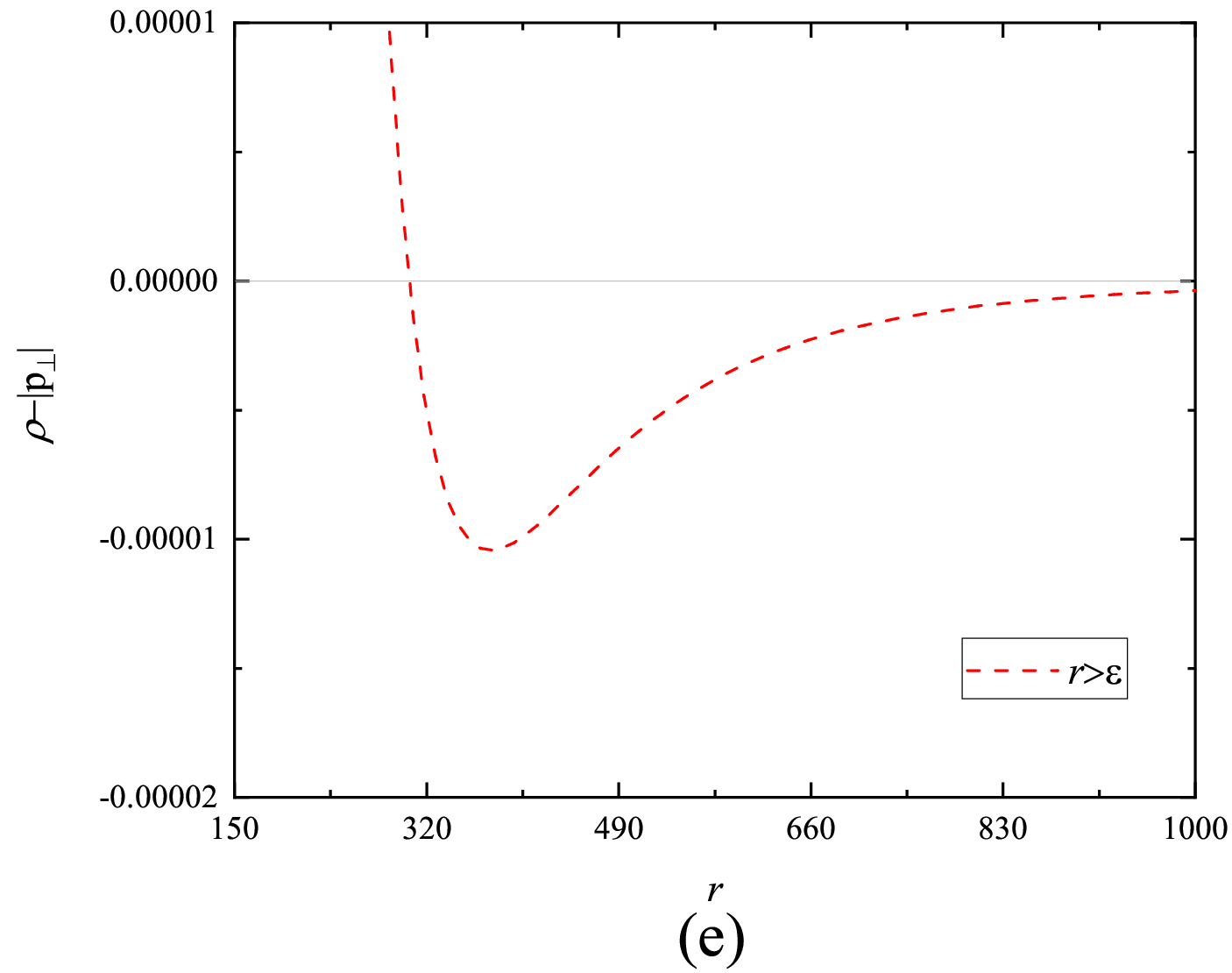}
\label{fig4-e+}
\end{minipage}
}
\subfigure{
\begin{minipage}{0.42\textwidth}
\vspace{-0.7cm}
\includegraphics[width=\textwidth]{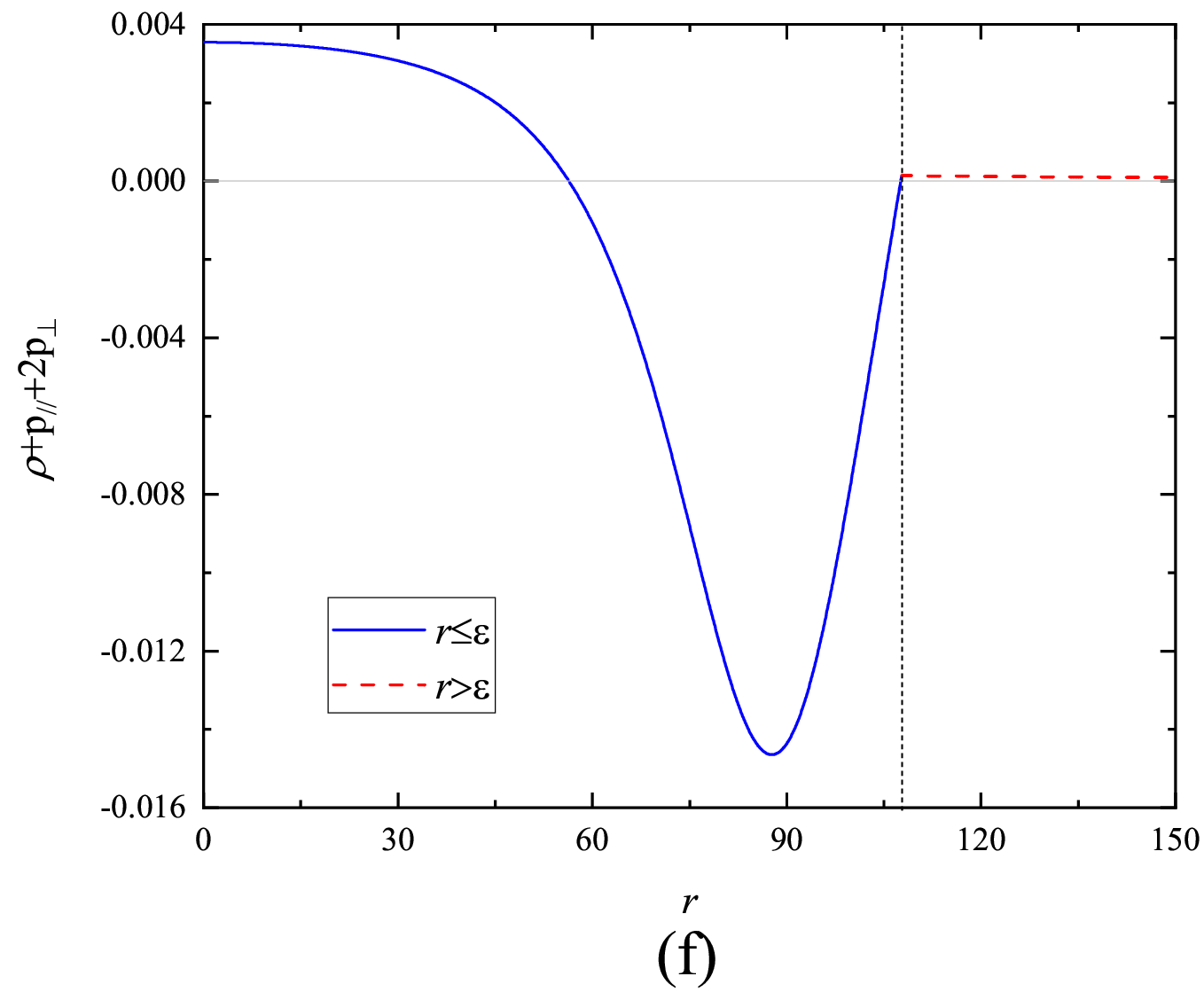}
\label{fig4-e}
\end{minipage}
}
\caption{Energy conditions as a function of $r$. (a) $\rho  + {p_{//}}$ versus $r$. (b) $\rho  + {p_ \bot }$ versus $r$. (c) $\rho  - \left| {{p_{//}}} \right|$ versus $r$. (d), (e) $\rho  - \left| {{p_{\bot}}} \right|$ versus $r$ for different radial intervals. (f) $\rho  +  {p_{//}} +2{p_ \bot }$ versus $r$. Here $a = 10 {m^{\frac{1}{3}}}$ and $m = 10^4$ (in Planck units).}
\label{fig4}
\end{figure}

\vspace{-\baselineskip}
\begin{align}
\label{eq15-1}
\rho  + {p_{//}} & =   \frac{{\left( {4{r^2} - {a^2}} \right)}\left( {11{a^8} - 244{a^6}{r^2} + 1568{a^4}{r^4} - 5376{a^2}{r^6} + 6144{r^8}} \right)}{{4\pi \mathcal{D}{{\left( {{a^6} - 11{a^4}{r^2} + 48{a^2}{r^4} - 64{r^6}} \right)}^2}{{\left( {\mathcal{D} - 2m} \right)}^2}}} \nonumber \\
 &   + \frac{{{a^2}\left( {\mathcal{D} - 6m} \right) + \mathcal{D}\left( {12{m^2} + {r^2}} \right) - 8{m^3} - 6m{r^2}}}{{4\pi \mathcal{D}{{\left( {{a^6} - 11{a^4}{r^2} + 48{a^2}{r^4} - 64{r^6}} \right)}^2}{{\left( {\mathcal{D} - 2m} \right)}^2}}},
 \end{align}
 \begin{align}
\label{eq15-2}
\rho  + {p_ \bot } & =  - \frac{{5{a^{10}} - 78{a^8}{r^2} + 192{a^6}{r^4} + 928{a^4}{r^6} - 4608{a^2}{r^8} + 6144{r^{10}}}}{{4\pi {{\left( {{a^6} - 11{a^4}{r^2} + 48{a^2}{r^4} - 64{r^6}} \right)}^2}}}
 \nonumber \\
& +\frac{{m\left( {23{a^{14}} - 314{a^{12}}{r^2} + 479{a^{10}}{r^4} + 3360{a^8}{r^6} - 4320{a^6}{r^8}} \right)}}{{8\pi {\mathcal{D}^5}{{\left( {{a^6} - 11{a^4}{r^2} + 48{a^2}{r^4} - 64{r^6}} \right)}^2}}}
 \nonumber \\
&+ \frac{{m\left( { - 22272{a^4}{r^{10}} + 47104{a^2}{r^{12}} + 16384{r^{14}}} \right)}}{{8\pi {\mathcal{D}^5}{{\left( {{a^6} - 11{a^4}{r^2} + 48{a^2}{r^4} - 64{r^6}} \right)}^2}}},
\end{align}
where  $\mathcal{D} = \sqrt {{a^2} + {r^2}}$. We numerically plot these quantities in Fig.~\ref{fig4-a} and Fig.~\ref{fig4-b} based on Eqs.~(\ref{eq13})-(\ref{eq15-2}).

In all these figures, the quantities inside the gluing surface are plotted with blue curves while the quantities outside the gluing surface are plotted with red curves.
It is obvious that all the red curves satisfy the condition of being greater than or equal to zero. In Fig.~\ref{fig4-a} and Fig.~\ref{fig4-b}, for $r \leq \varepsilon$ the blue curves initially decrease below zero ($a_1$ and $b_1$) and then rise above zero ($a_2$ and $b_2$) as they approach $r = 0$ due to the effects of QG. The above results demonstrate that the NEC is violated only in a small region of the spacetime, which greatly improves the situation in the original spacetime without modification.

In addition, from Fig.~\ref{fig3} and Fig.~\ref{fig4}, one notices that the WEC and SEC are violated only in a confined region inside the gluing surface as well, mirroring the behavior observed in the violation of the NEC. This violation in a finite region highlights how the influence of QG effectively halts the collapse at the center of the black hole, thereby opening up the possibility for matter to traverse through it. Nevertheless, we point out that the DEC is an exception, since it drops below zero outside the gluing surface and converges to zero at infinity (see Fig.~\ref{fig4-e+}). This is due to the property of the black hole spacetime itself. In the next section, we apply a more effective modification scheme that may improve the situation.
\vspace{-0.7cm}

\section{New black-to-white hole solutions II}
\label{sec3}
In this section we present another interesting black-to-white hole solution based on standard Schwarzschild black hole. We will introduce a gluing surface inside the horizon such that the metric for the region inside the gluing surface will be modified, while the geometry outside the gluing surface remains identical with Schwarzschild geometry. Therefore, the energy conditions are automatically satisfied outside the gluing surface.
\vspace{-0.5cm}
\subsection{The metric form of the new black-to-white solution}
\label{sec3-1}
\vspace{-0.1cm}
Inspired by Ref.~\cite {cha27}, we start with the metric for the interior of the horizon in Schwarzschild spacetime, and then introduce a gluing surface at $\tau=\varepsilon$ and modify the metric as
\begin{equation}
\label{eq17}
{\text{d}}{s^2} =  - \frac{{4{g^2}\left( {\tau ,l} \right)}}{{2m - {\tau ^2}}}{\text{d}}{\tau ^2} + \frac{{2m - {\tau ^2}}}{{g \left( {\tau ,l} \right)}}{\text{d}}{x^2} + {g^2}\left( {\tau ,l} \right){\text{d}}{\Omega ^2},
\end{equation}
with
\begin{subequations}
\label{eq17+}
\begin{numcases}{{g}\left( {\tau ,l} \right) =}
{\tau^2} + {l}{\left( {1 - \frac{{{\tau^2}}}{{{\varepsilon ^2}}}} \right)^n}, & for ${\tau   \leq \varepsilon }$,\\
{\tau^2}, & for ${\tau > \varepsilon}$,
\end{numcases}
\end{subequations}
where $\tau$ and  $x$ are time and spatial coordinates inside the horizon, respectively. They relate to the ordinary coordinates in Schwarzschild metric by $t=x$ and $r=\tau^2$.  It is easy to check that for $\tau > \varepsilon$, the metric is nothing but the metric of Schwarzschild spacetime, given by ${\text{d}}{s^2} = - \left( {1 - {{2m} \mathord{\left/ {\vphantom {{2m} r}} \right. \kern-\nulldelimiterspace} r}} \right){\text{d}}{t^2} + {\left( {1 - {{2m} \mathord{\left/ {\vphantom {{2m} r}} \right. \kern-\nulldelimiterspace} r}} \right)^{ - 1}}{\text{d}}{r^2} + {r^2}{\text{d}}{\Omega ^2}$.  For the same consideration in Section~\ref{sec2}, to ensure the continuity of the Kretschmann scalar curvature as well as the stress-energy tensor at the gluing surface, we consider the case with $n = 3$. In addition, we remark that  in contrast to the situation described in Section~\ref{sec2}, the components of ${\text{d}}{\tau ^2}$ and ${\text{d}}{x ^2}$ in this metric are no longer inverse to each other.

\begin{figure}[H]
\centering
\includegraphics[width=0.55\textwidth]{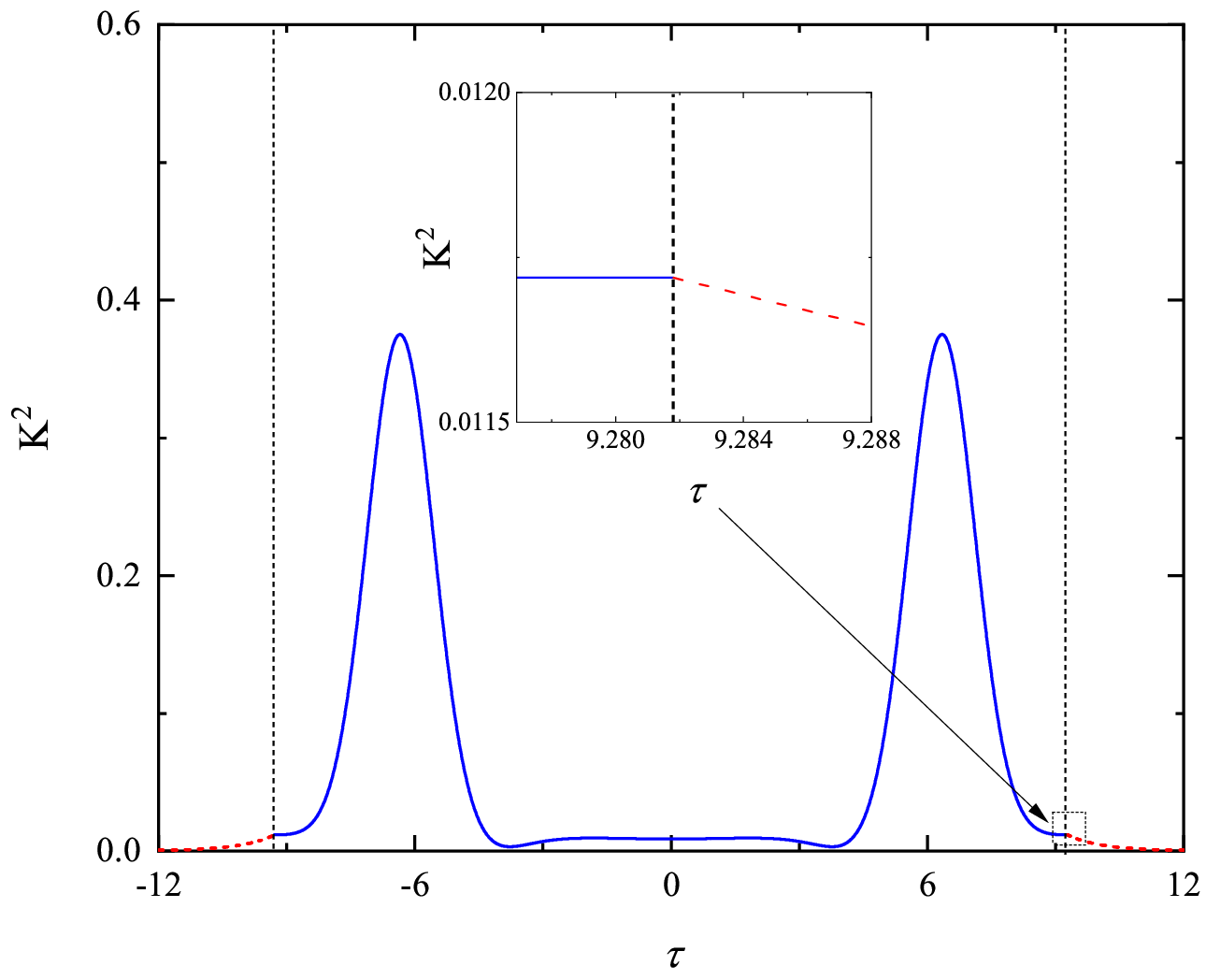}
\caption{The Kretschmann scalar curvature  $K^2$  as a function of coordinate $\tau$ for $l = 4 {m^{\frac{1}{3}}}$ and $m = 10^4$ (in Planck units).}
\label{fig5}
\end{figure}

Now it is straightforward to calculate the Kretschmann scalar curvature, whose forms inside and outside the gluing surfaces are presented in Appendix~\ref{appD}. In particular, for $\tau=0$, one has
\begin{equation}
\label{eq19+}
{K^2}\left( {\tau  = 0} \right) = \frac{{9{m^2}}}{{{l^6}}} + \frac{{81{m^2}}}{{{l^4}{\varepsilon ^4}}} + \frac{{17}}{{4{l^4}}} + \frac{m}{{{l^5}}} - \frac{{3m}}{{{l^4}{\varepsilon ^2}}} - \frac{{54{m^2}}}{{{l^5}{\varepsilon ^2}}}.
\end{equation}
As we discussed in the previous section, to keep the curvature below the Planck scale independent of the mass of black hole, we may set the value of the parameter $l\sim m^{\frac{1}{3}}$. Without loss of generality we set $l = 4 {m^{\frac{1}{3}}}$. Similarly, we specify the position of the gluing surface by taking $\varepsilon  = \sqrt l $,  then the Kretschmann scalar curvature takes some simple form inside and outside the gluing surfaces, as plotted in Fig.~\ref{fig5}.

In Fig.~\ref{fig5}, the value of Kretschmann scalar curvature is plotted with blue curve inside the gluing surface, while with red curve outside the gluing surface, whose location is marked by the vertical dashed line. We notice that Kretschmann scalar curvature is continuous at the gluing surface, which is also zoomed in the inset. Furthermore, when $\tau$ reaches zero, Kretschmann scalar curvature attains a finite value, indicating the global regularity of the geometry described by Eq.~(\ref{eq17}). It is also noteworthy that its maximal value is below the Planck scale, but its position shifts from the center of the black hole to some position between the gluing surface and the core. In addition, its mass independence is justified by setting a very large value for the mass of black hole.

\begin{figure}[H]
\centering
\includegraphics[width=0.7\textwidth]{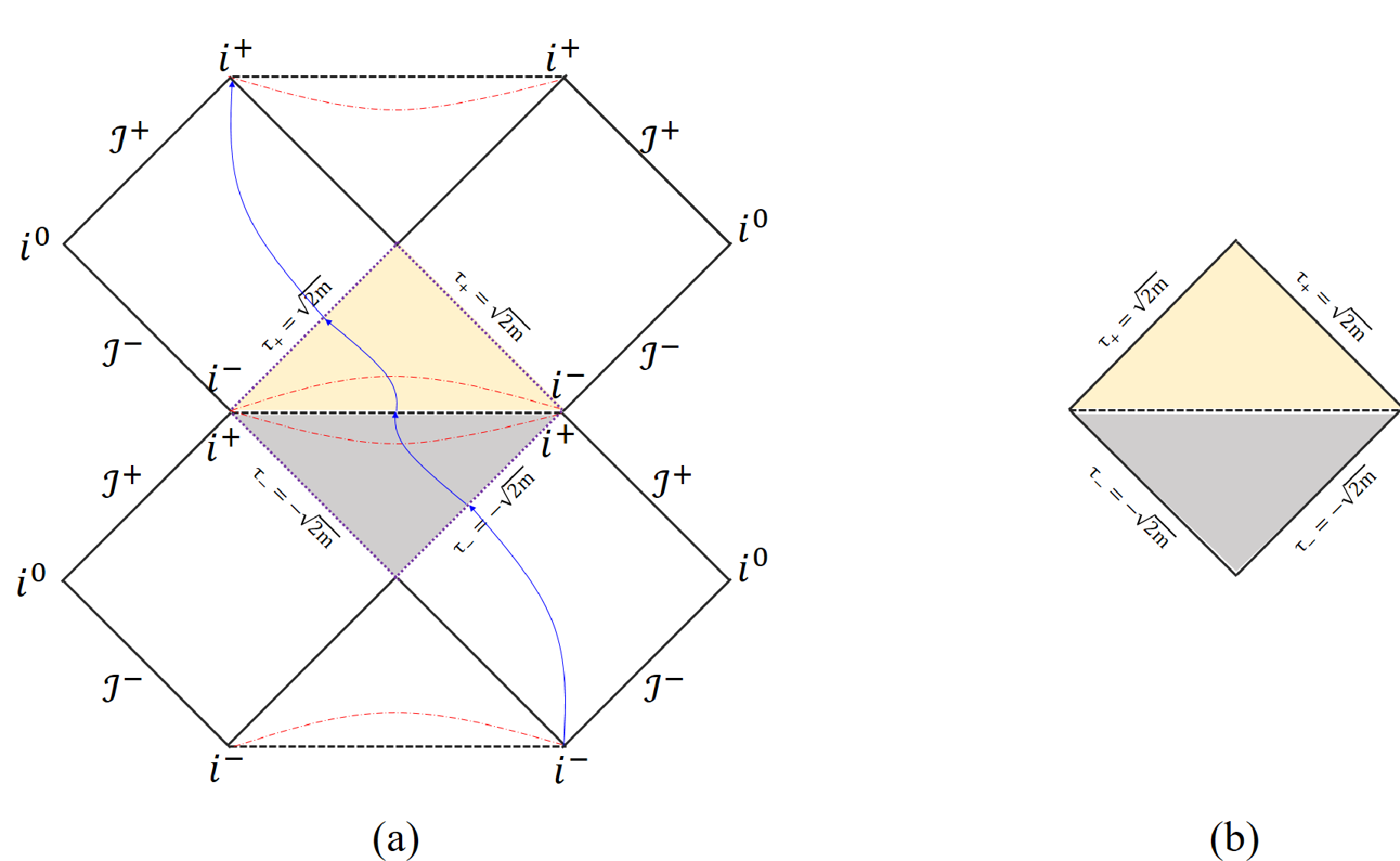}
\caption{(a) Carter-Penrose diagram for the maximal spacetime extension of line element~(\ref{eq17}). (b) Carter-Penrose diagram for the spacetime with line element~(\ref{eq1}).}
\label{fig8}
\end{figure}

Next we briefly discuss the geodesics of particles near the core of the black hole. In this case, the radial geodesic equation is  ${g_{xx}}{\left( {{{{\text{d}}x} \mathord{\left/ {\vphantom {{{\text{d}}x} {{\text{d}}\chi }}} \right. \kern-\nulldelimiterspace} {{\text{d}}\chi }}} \right)^2} + {g_{\tau \tau }}{\left( {{{{\text{d}}\tau } \mathord{\left/ {\vphantom {{{\text{d}}\tau } {{\text{d}}chi }}} \right. \kern-\nulldelimiterspace} {{\text{d}}\chi }}} \right)^2} =  - \delta $.  Taking the conserved quantity into account, i.e., $E =  - {g_{xx}}\left( {{{{\text{d}}t}\mathord{\left/ {\vphantom {{{\text{d}}\tau} {{\text{d}}\chi }}} \right. \kern-\nulldelimiterspace} {{\text{d}}\chi }}} \right)$, the affine parameter for the integration with respect to the radial coordinate $\tau$ can be expressed as
\begin{equation}
\label{eq19+}
\Delta \chi  = \int_{{\tau_p}}^{{\tau_q}} {\sqrt { - \frac{{{g_{xx}}{g_{\tau \tau }}}}{{\delta {g_{xx}} + {E^2}}}} {\text{d}}\tau },
\end{equation}
with the initial moment of motion $\tau_{q}$ and the final moment $\tau_{p}$.  For massless particles (i.e., $\delta=0$), the affine parameter is
\begin{equation}
\label{eq20+1}
\Delta \chi \left( {\tau  > \varepsilon } \right) = \frac{2}{E }\int_{{\tau _p}}^{{\tau _q}} \tau {\text{d}}\tau,
\end{equation}
\begin{equation}
\label{eq20+2}
\Delta \chi \left( {\tau  \leq \varepsilon } \right)  = \frac{2}{E }\int_{{\tau _p}}^{{\tau _q}} {\sqrt {{\tau ^2} + l{{\left( {1 - \frac{{{\tau ^2}}}{{{\varepsilon ^2}}}} \right)}^3}} } {\text{d}}\tau.
\end{equation}
For massive particles (i.e., $\delta=1$), the affine parameter can be expressed as
\begin{equation}
\label{eq21+1}
\Delta \chi  \left( {\tau >  \varepsilon } \right) = \int_{{\tau _p}}^{{\tau _q}} {\frac{{2\tau }}{{\sqrt {{{2m} \mathord{\left/
 {\vphantom {{2m} {{\tau ^2}}}} \right. \kern-\nulldelimiterspace} {{\tau ^2}}} + {E ^2} - 1} }}} {\text{d}}\tau,
 \end{equation}
 \begin{equation}
\label{eq21+2}
\Delta \chi  \left( {\tau \leq   \varepsilon } \right)  = \int_{{\tau _p}}^{{\tau _q}} {\frac{{2\left[ {{\varepsilon ^6}{\tau ^2} + l{{\left( {{\varepsilon ^2} - {\tau ^2}} \right)}^3}} \right]}}{{{\varepsilon ^3}\sqrt {{\varepsilon ^6}\left[ {2m + l{E ^2} + {\tau ^2}\left( {{E^2} - 1} \right)} \right] - l{\tau ^2}\left( {3{\varepsilon ^4} - 3{\varepsilon ^2}{\tau ^2} + {\tau ^4}} \right){E ^2}} }}} {\text{d}}\tau.
\end{equation}
From Eqs.~(\ref{eq20+1})-(\ref{eq21+2}), it is found that the integrands are always finite if the mass $m$ and the energy $E$ are bounded  (for the sake of simplicity, one can set $\varepsilon=\sqrt{l}$ and $l=4 m^{1/3}$) and the integration interval is also finite. Thus, the results of the integration should be finite as well, which indicates that spacetime is not complete in the region $ 0  < \tau  <  + \infty$, instead it should be extended to $ - \infty < \tau < + \infty$. Now, we would like to further analyze the  structure of the spacetime after extension. According to Eq.~(\ref{eq17}), the Penrose diagram of the maximal spacetime extension is illustrated in tin Fig.~\ref{fig8}(a). First of all, since we introduce the gluing surface at $\tau=\varepsilon$, which is marked by dotted-dashed red line in figure, and remove the singularity at $\tau=0$, we find the classical Kruskar diagram can be extended and form a new Penrose diagram for a black-to-white hole. The way of construction indicates that the effect of QG is confined within the gluing surface, while outside this region, the spacetime reduces to the familiar Schwarzschild black hole.  As a result, our proposed line element successfully describes the spacetime structure within the range $\tau \in \left( { - \infty ,\infty } \right)$, which may be viewed as the extension of metric in Eq.~(\ref{eq1}) that was originally presented in \cite{chb27} and limited to describe the interior of the horizon, as illustrated in Fig.~\ref{fig8}(b). Secondly, a similar analysis shows that the geodesics of particles may travel from the black hole region (grey diamond), passing through the core, to the white hole region (yellow diamond), as illustrated as the blue line in the diagram.

\subsection{The stress-energy tensor and energy conditions}
\label{sec3-2}
\begin{figure}[htbp]
\setlength{\abovecaptionskip}{-0.4cm}
\setlength{\belowcaptionskip}{-0.1cm}
\subfigure{
\begin{minipage}{0.42\textwidth}
\centering
\includegraphics[width=\textwidth]{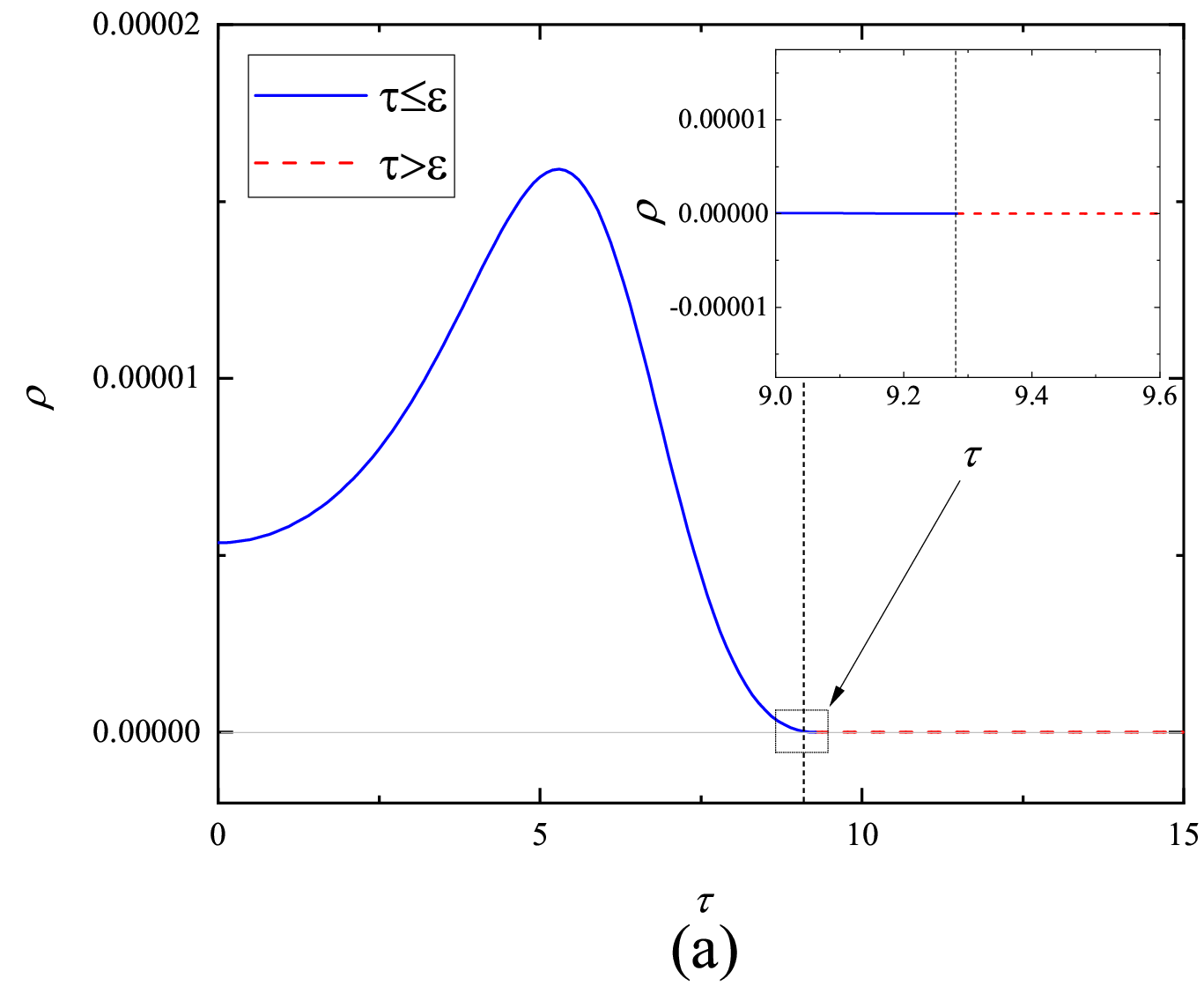}
\label{fig6-a}
\end{minipage}
}
\subfigure{
\begin{minipage}{0.42\textwidth}
\centering
\includegraphics[width=\textwidth]{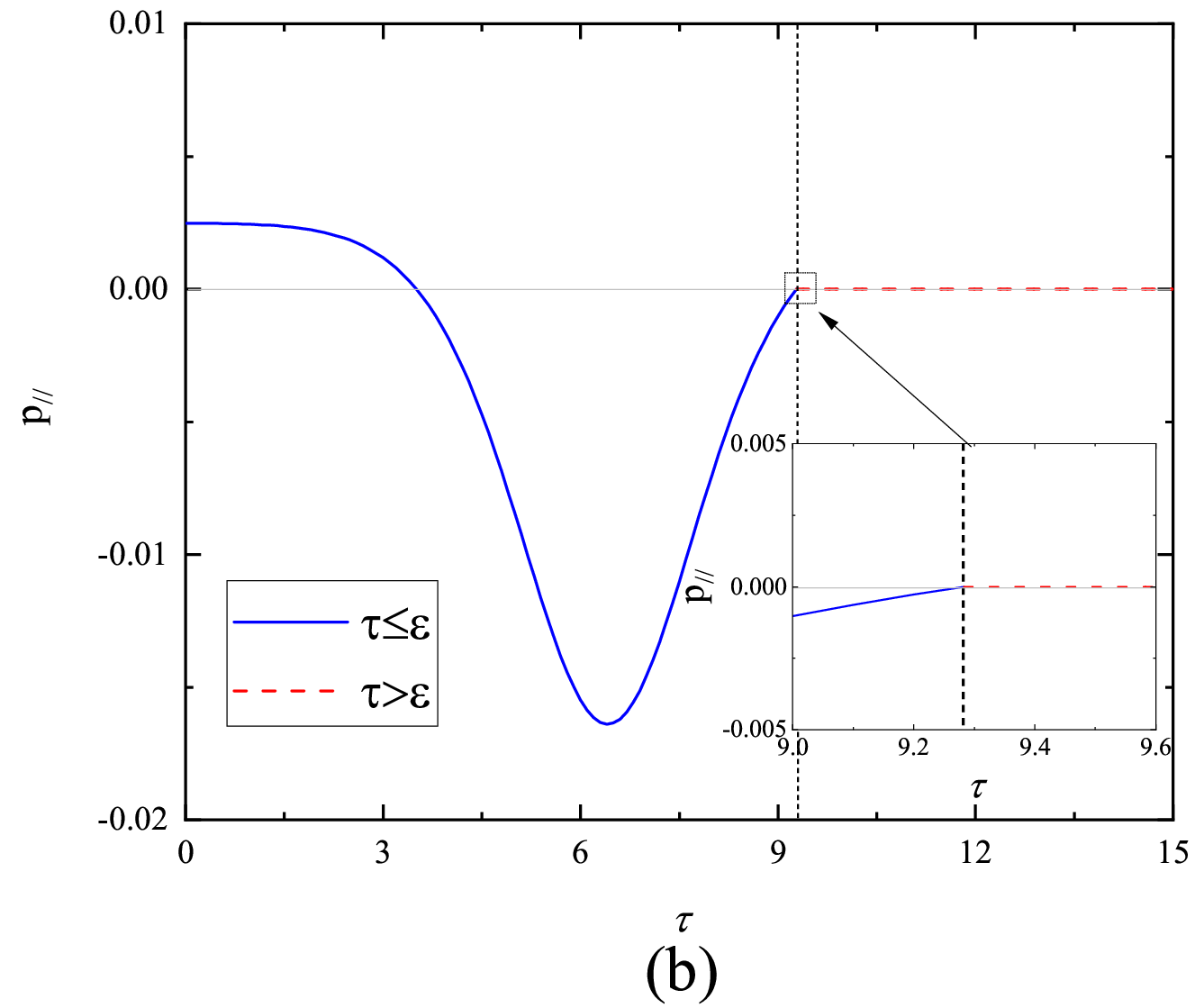}
\label{fig6-b}
\end{minipage}
}
\subfigure{
\begin{minipage}{0.42\textwidth}
\vspace{-0.5cm}
\centering
\includegraphics[width=\textwidth]{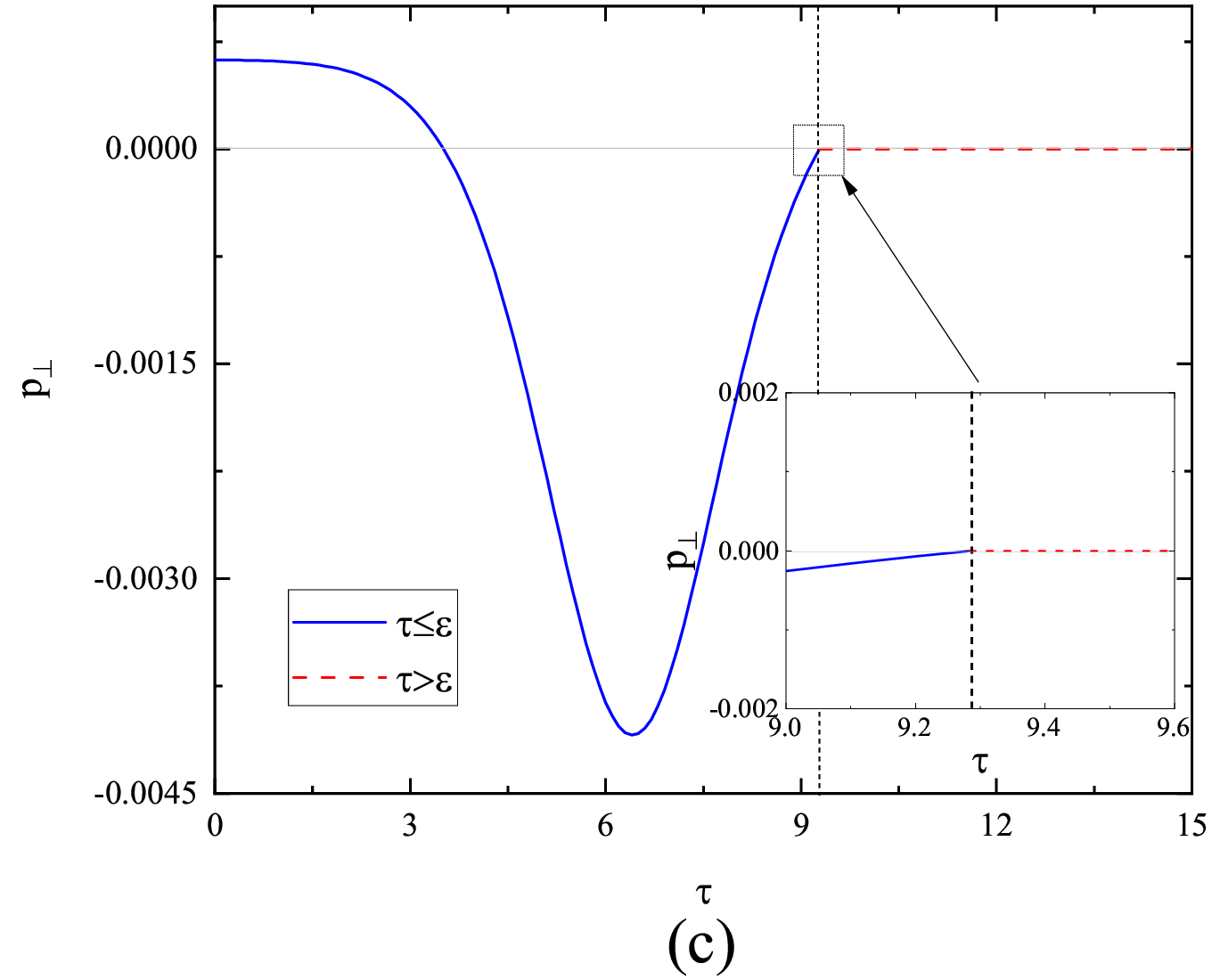}
\label{6-c}
\end{minipage}
}
\caption{The relationship between energy density and pressure with $\tau$. (a) $\rho $ versus $\tau$. (b) $ {p_ {//} }$ versus $\tau$. (c) ${p_ {\bot} }$ versus $\tau$.
Here $l = 4 {m^{\frac{1}{3}}}$ and $m = 10^4$ (in Planck units).}
\label{fig6}
\end{figure}

Now, we turn to the energy conditions in this spacetime. Since outside the gluing surface it is the standard Schwarzschild solution with $\rho  ={p_{//}} = {p_ \bot }= 0$, we only need to consider the energy conditions inside the gluing surface. According to Eqs.~(\ref{eq7-1})-(\ref{eq7-3}), the non-zero components of stress-energy tensor~(\ref{eq17}) within the region $r \leq \varepsilon$ are given by
\begin{equation}
\label{eq20-1}
T_\tau ^\tau  =  - \frac{{l{\varepsilon ^{12}}\left( {{\varepsilon ^6} - 3{\varepsilon ^2}{\tau ^4} + 2{\tau ^6}} \right)}}{{8\pi{{\left( {{\varepsilon ^6}{\tau ^2} + l {\mathcal{X}^3}} \right)}^3}}},
\end{equation}
\begin{align}
\label{eq20-2}
T_x^x & =  - \frac{{l{\varepsilon ^{12}}\mathcal{X}}}{{8\pi {{\left( {{\varepsilon ^6}{\tau ^2} + l{\mathcal{X}^3}} \right)}^4}}}\left\{ {l{\mathcal{X}^3}\left[ {{\varepsilon ^4} - 6m{\varepsilon ^2} + 4\left( {3m + {\varepsilon ^2}} \right){\tau ^2} - 8{\tau ^4}} \right]} \right.
\nonumber \\
&\left. { + 2{\varepsilon ^6}\left[ {m\left( {{\varepsilon ^4} + {\varepsilon ^2}{\tau ^2} + 10{\tau ^4}} \right) - 6{\tau ^6}} \right]} \right\},
\end{align}
 \begin{align}
\label{eq20-3}
T_\theta ^\theta &  = T_\phi ^\phi  = \frac{{l{\varepsilon ^{12}}\mathcal{X}}}{{32\pi {{\left( {{\varepsilon ^6}{\tau ^2} + l{\mathcal{X}^3}} \right)}^4}}}\left\{ {l{\mathcal{X}^3}\left[ {6m{\varepsilon ^2} + {\varepsilon ^4} - 2\left( {6m + {\varepsilon ^2}} \right){\tau ^2} + 4{\tau ^4}} \right]} \right.
\nonumber \\
&+ \left. {2{\varepsilon ^6}\left[ {{\varepsilon ^2}\left( {{\varepsilon ^2} - m} \right){\tau ^2} - m{\varepsilon ^4} + \left( {{\varepsilon ^2} - 10m} \right){\tau ^4} + 4{\tau ^6}} \right]} \right\},
\end{align}
where $\mathcal{X}={\varepsilon ^2} - {\tau ^2}$ and the corresponding energy density and pressure are $\rho  = T_\tau ^\tau$ , ${p_{//}} = T_x^x$, ${p_ \bot } = T_\theta ^\theta  = T_\phi ^\phi$. We plot these quantities as a function of $\tau$ in Fig.~\ref{fig6}. One can see that all these quantities  are continuous at the gluing surface (vertical dashed line),  indicating that the spacetime geometry~(\ref{eq17}) is continuous as a whole. Interestingly, Fig.~\ref{fig6-a} shows that the energy density is always lager than zero inside the gluing surface, and this behavior can also be obtained by analyzing Eq.~(\ref{eq20-1}), i.e., $\rho = 0$ only at $\tau = \varepsilon$.

Then, let us check the NEC. Using Eqs.~(\ref{eq20-1})-(\ref{eq20-3}), one has
\begin{equation}
\label{eq21-1}
\rho  + {p_{//}} = \frac{{{l^4}\left( {2m - {\tau ^2}} \right)\left( {2{l^5} - 18{l^4}{\tau ^2} + 33{l^3}{\tau ^4} - 38{l^2}{\tau ^6} + 27l{\tau ^8} - 6{\tau ^{10}}} \right)}}{{8\pi{{\left( {{l^3} - 2{l^2}{\tau ^2} + 3l{\tau ^4} - {\tau ^6}} \right)}^4}}},
\end{equation}
\begin{align}
\label{eq21-2}
\rho  + {p_ \bot } & = \frac{{{l^4}\left( {l - {\tau ^2}} \right)}}{{32\pi {{\left( {{l^3} - 2{l^2}{\tau ^2} + 3l{\tau ^4} - {\tau ^6}} \right)}^4}}}\left[ {5{l^5} + {l^4}\left( {4m - 7{\tau ^2}} \right) + {l^3}\left( {11{\tau ^4}} \right.} \right.
\nonumber \\
& \left. {\left. { - 32m{\tau ^2}} \right) + {l^2}\left( {34m{\tau ^4} + 13{\tau ^6}} \right) - 14l\left( {3m{\tau ^6} + {\tau ^8}} \right) + 4{\tau ^8}\left( {3m + {\tau ^2}} \right)} \right].
\end{align}

\begin{figure}[htbp]
\centering
\subfigure{
\begin{minipage}{0.42\textwidth}
\includegraphics[width=\textwidth]{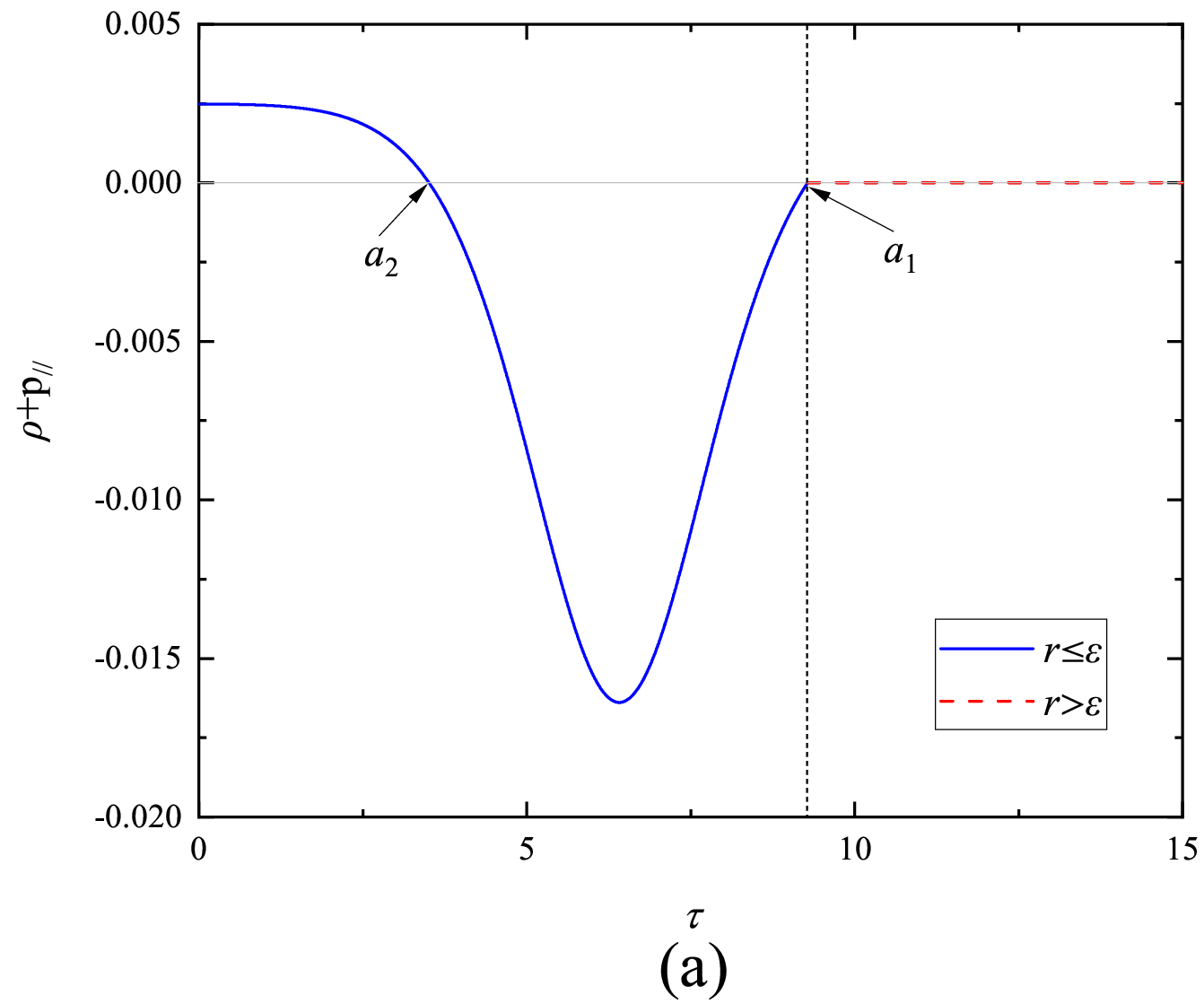}
\label{fig7-a}
\end{minipage}
}
\subfigure{
\begin{minipage}{0.42\textwidth}
\includegraphics[width=\textwidth]{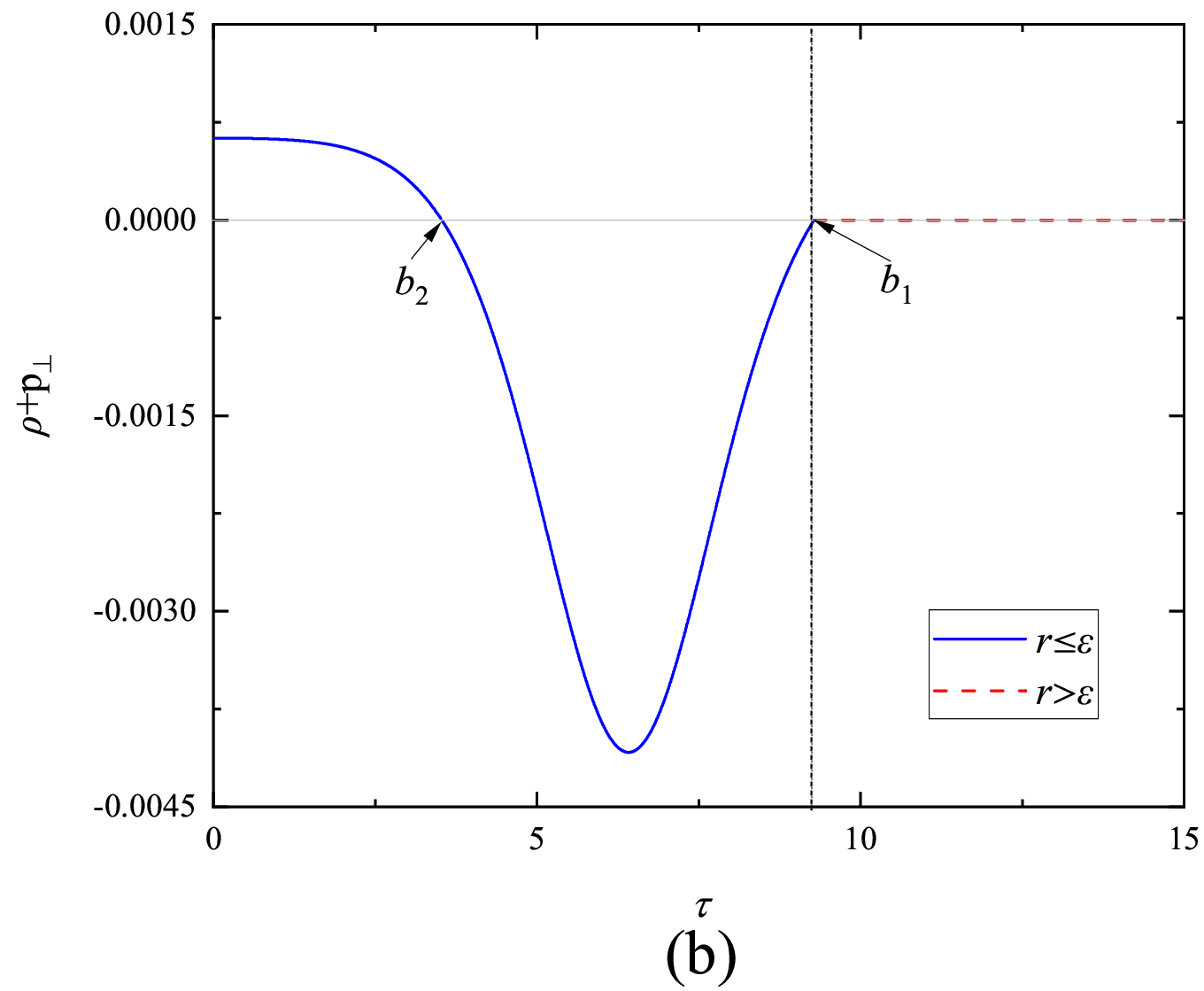}
\label{fig7-b}
\end{minipage}
}
\subfigure{
\begin{minipage}{0.42\textwidth}
\vspace{-0.5cm}
\includegraphics[width=\textwidth]{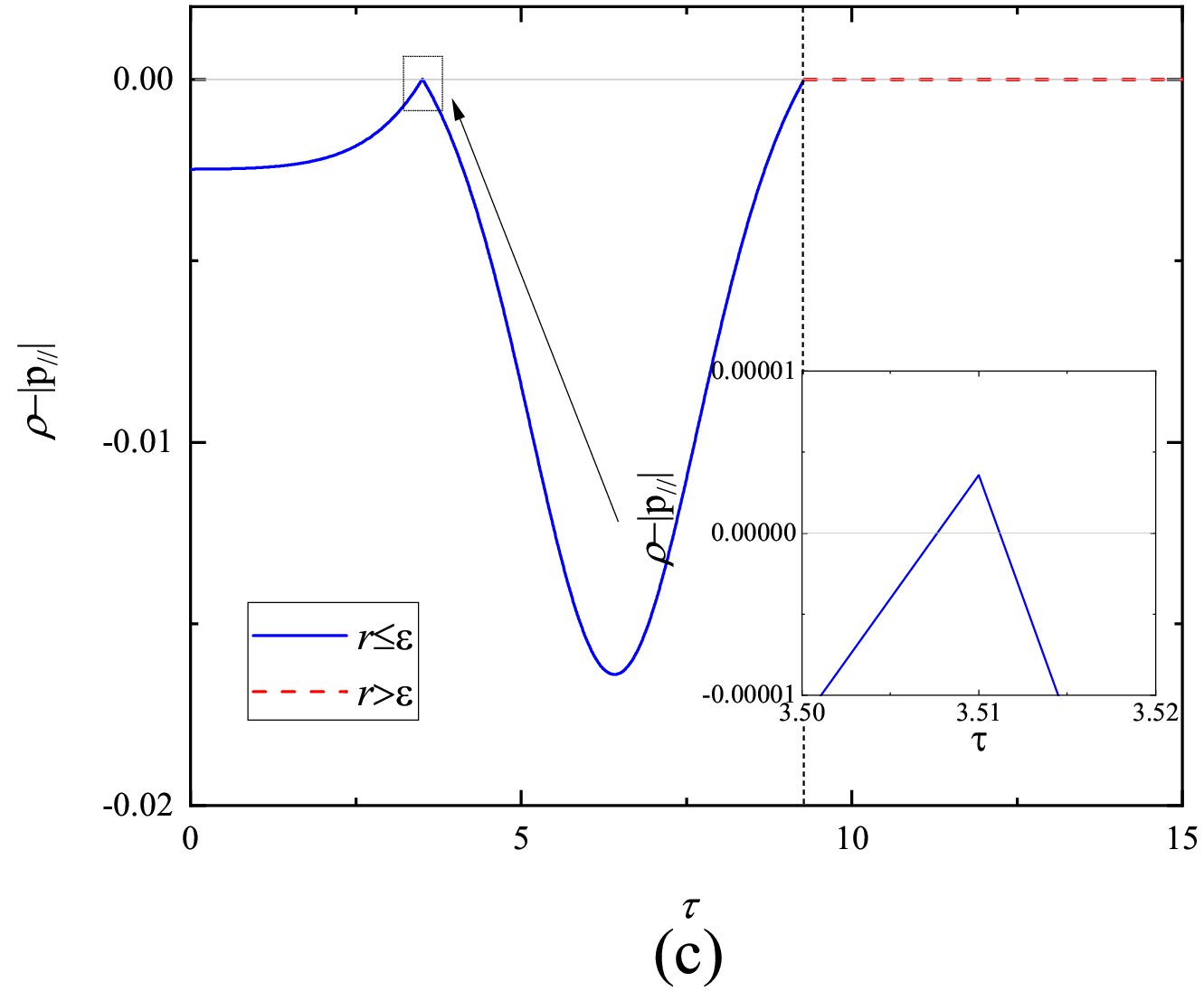}
\label{fig7-c}
\end{minipage}
}
\subfigure{
\begin{minipage}{0.42\textwidth}
\vspace{-0.5cm}
\includegraphics[width=\textwidth]{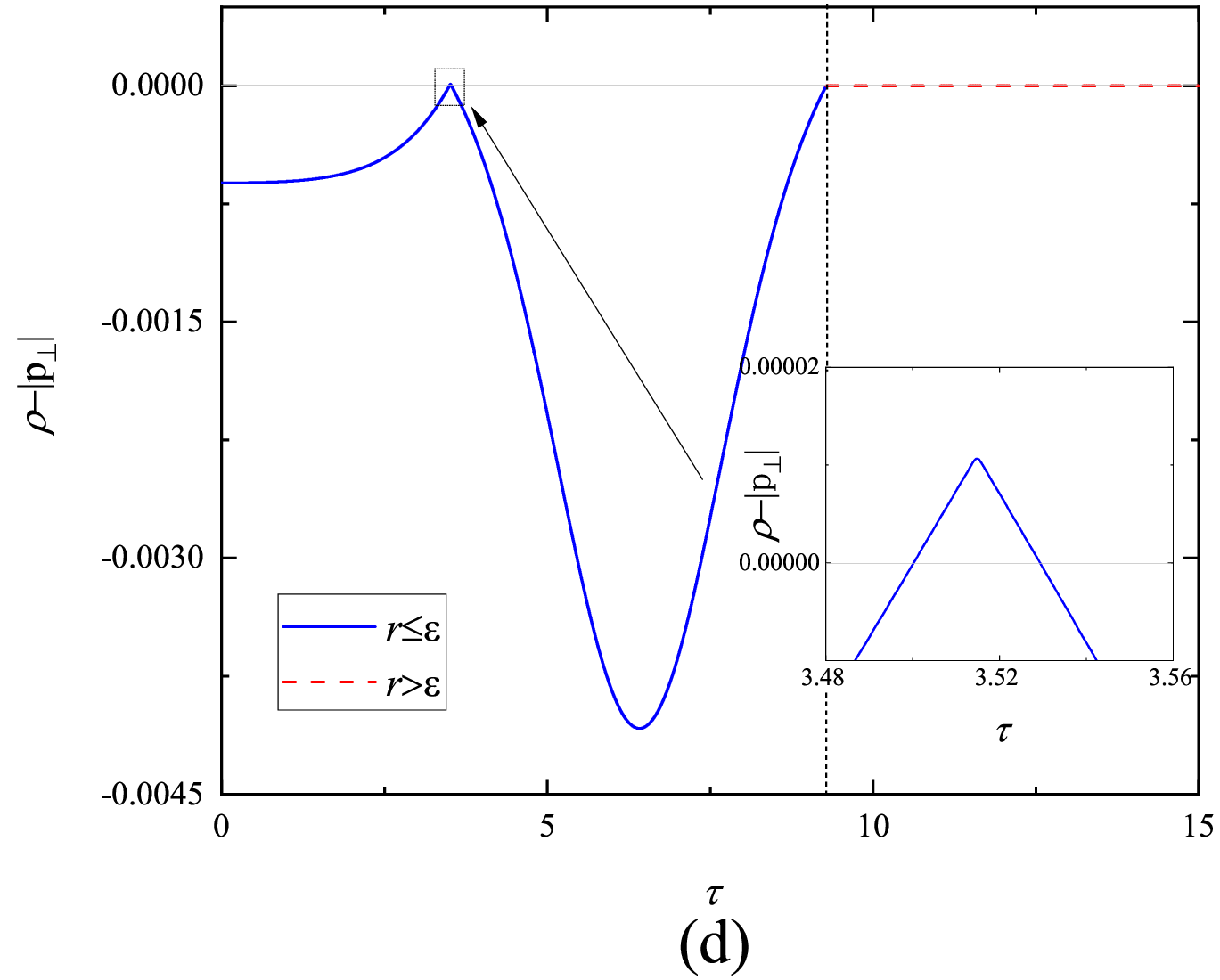}
\label{fig7-d}
\end{minipage}
}
\subfigure{
\begin{minipage}{0.42\textwidth}
\vspace{-0.5cm}
\includegraphics[width=\textwidth]{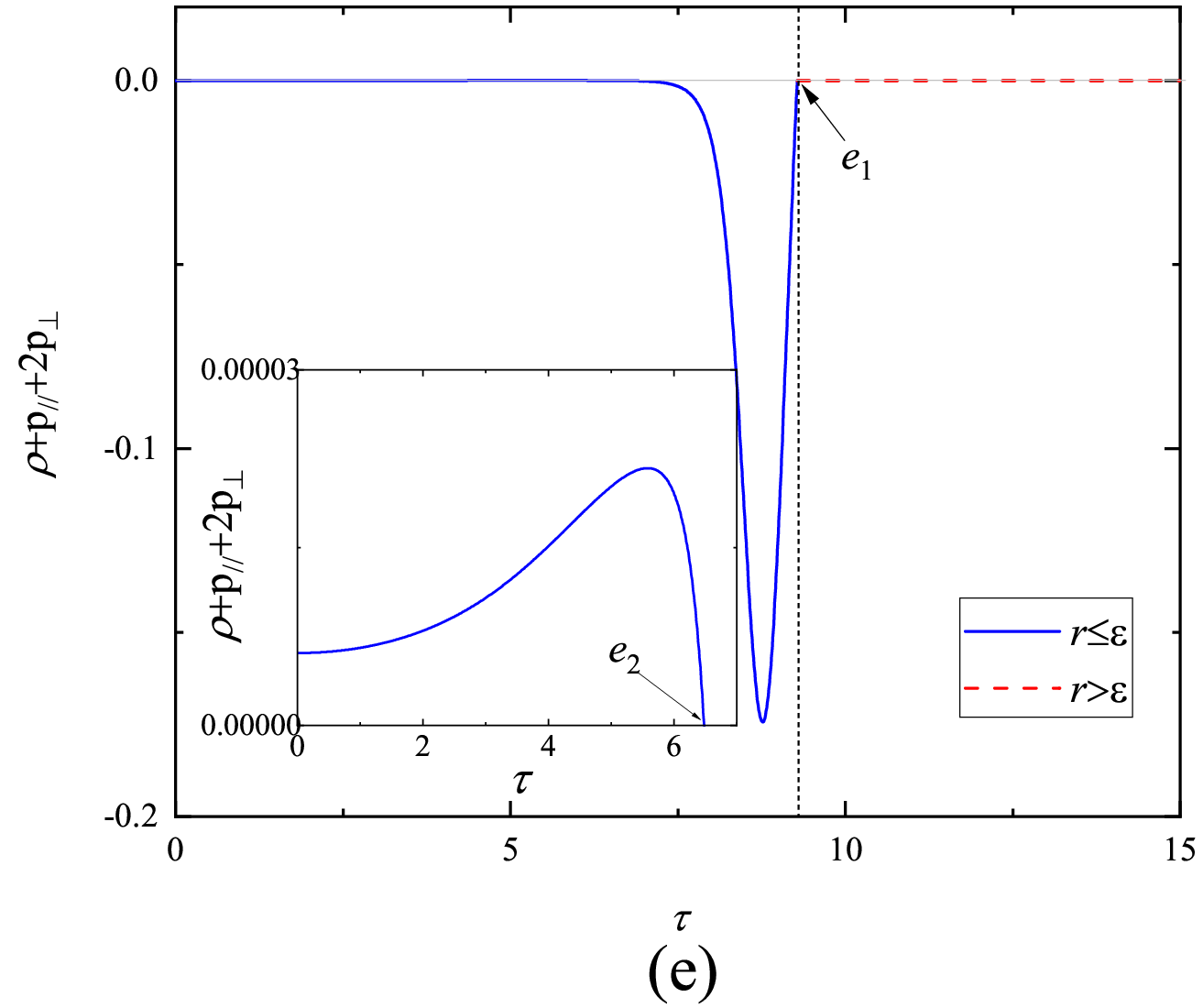}
\label{fig7-e}
\end{minipage}
}
\caption{Energy conditions as a function of $\tau$. (a) $\rho  + {p_{//}}$ versus $\tau$. (b) $\rho  + {p_ \bot }$ versus $\tau$.
 (c) $\rho  - \left| {{p_{//}}} \right|$ versus $\tau$. (d) $\rho  - \left| {{p_{\bot}}} \right|$ versus $\tau$. (e) $\rho  +  {p_{//}} +2{p_ \bot }$ versus $\tau$.
 Here $l = 4 {m^{\frac{1}{3}}}$ and $m = 10^4$ (in Planck units).}
\label{fig7}
\end{figure}

Fig.~\ref  {fig7-a} and Fig.~\ref  {fig7-b} illustrate  the behavior of  $\rho  + {p_{//}}$ and $\rho  + {p_ \bot }$ as the function of $\tau$.  Without surprise, the NEC is always satisfied outside the gluing surface, while inside the gluing surface  the NEC is voilated only  within a small region (i.e., $\left( {{a_1},{a_1}} \right)$ and $\left( {{b_2},{b_2}} \right)$).

In the end of this section we present a brief remark about the other energy conditions in Table~\ref{tab1}. It becomes evident that the SEC, DEC, and WEC are also violated solely in a minute region within the gluing surface (as depicted in Fig.~\ref{fig6}-Fig.~\ref{fig7}). Consequently, we can deduce that the effect of quantum gravity results in the violation of the energy conditions only in a small region close to the core of black hole, which in turn prevents the occurrence of a singularity and establishes a pathway to reach the white hole. Recent a lot of relevant work have explored the possibility of revealing the black-to-white hole structure by photon rings, for instance, in Ref.~\cite{chx35}. Definitely, the black-to-white hole solutions found in current paper  provide new arena for exploring these fascinating pictures. It is very worthwhile to investigate the related problems on observational signatures
separately in future.

\section{Conclusions}
\label{sec4}
In this paper we have proposed a simple strategy to construct new black-to-white solutions with spherical symmetry, confining the effect of QG to a localized gluing surface region near the core of the black hole such that the energy conditions are obeyed in most region of the spacetime.

In the first solution with metric (\ref{eq10}), the strategy of construction solely affects the angular component of the metric ${g^2}\left( {r,a} \right)$. Remarkably, an intriguing observation is that the violation of the NEC is confined to a small region within the truncation point, thus presenting a notable improvement over the S-V solution, where NEC violations are pervasive throughout the entire spacetime. In the second solution with metric (\ref{eq17}), our strategy of construction extends to all components of the metric for the interior of Schwarzschild black hole. The new line element ensures a continuous connection between the interior and exterior of the event horizon, enabling the extension of the spacetime to encompass the range $\tau \in \left( { - \infty ,\infty } \right)$. Moreover, in both solutions the resulting effect of QG eliminates the singularity at $r=0$, allowing for the passage of matter from the black hole region in our universe to the white hole region in another universe. Consequently, this approach provides fascinating ideas for understanding the black hole information loss paradox.

Throughout this paper we have only considered the specific modification with $n=3$ in Eq.~(\ref{eq9}) and Eq.~(\ref{eq17+}), which guarantees the continuity of spacetime curvature and the energy-stress tensor of matter at the gluing surface. As a matter of fact, one may improve the continuity of the metric at the gluing surface up to the arbitrary order with a larger number $n$. In particular, as $n$ goes to infinity, the geometry of spacetime becomes smooth at the gluing surface. Nevertheless, we point out that as $n$ increases, one needs to choose a larger constant number $a_0$ to maintain Kretschmann scalar curvature to be sub-Planckian.

Finally, we intend to address the issue on the theoretical foundations which could lead to such new black-and-white holes. It is an important issue if one could obtain these solutions by solving any classical gravitational field equations coupled to matter fields, or the semi-classical effective equations due to the effects of QG. Currently, there are two primary approaches to explore the theoretical underpinnings of black-to-white holes. One approach involves classical theories with exotic matter, such as nonlinear electrodynamics~\cite{chyy1,chyy2,chyy3,chyy4}. The other approach considers the formation of black-to-white holes through the modification of QG~\cite{chr25,chr26,chr27,chr28,chr29,chr30,chr31,chr32,chr33,chr34,chr35,chr36,chr37,chr38}. In the former approach, Rodrigues and Silva have recently obtained the source of V-S spacetime and pointed out that V-S black hole is the solution to the Einstein's equation with known action of matter fields~\cite{chb32}. This progress also indicates that solutions constructed on the basis of V-S spacetime may be regarded as solutions of the gravity theory coupled to known matter  fields~\cite{chb33,chb34}. Encouraged by the above progress, we expect that our new solutions could also be understood as solutions to matter field equations, but we leave this issue for further investigation.In the latter approach, one candidate involves applying loop quantum gravity to the interior of black holes~\cite{chy3,chx34,chb35,chb36,chb37,chb38,chb39,chb40}. Remarkably, this approach has yielded black-to-white hole solutions with different forms, inspiring us to deepen our understanding of the origin of these black-to-white holes in the current work.

\acknowledgments
\vspace{-0.43cm}
This work is supported in part by the Natural Science Foundation of China (Grant Nos.~12035016,~12275275,~12105231 and~12275350). It is also supported by Beijing Natural Science Foundation (Grant No.~1222031), and by the Sichuan Science and Technology Program (Grant Nos. 2024NSFSC0456 and 2023NSFSC1348), the Sichuan Youth Science and Technology Innovation Research Team (Grant No.~21CXTD0038). We are very grateful to Guo-Ping Li, Pan Li, Yuxuan Liu, Yu Tian, Meng-HeWu, Zhangping Yu, Wei Zeng, Xia Zhou and Hongbao Zhang for helpful discussions.

\appendix
\setcounter{equation}{0}
\renewcommand\theequation{A.\arabic{equation}}
\section{The Kretschmann scalar curvature of solution I}
\label{appA}
The Kretschmann scalar curvature of Eq.~(\ref{eq10}) inside the gluing surface is given by
\begin{align}
\label{eqa1-1}
{K^2}\left( {r \leq \varepsilon } \right) & = \frac{2}{{{{\left( {{a^2} + {r^2}} \right)}^5}{\mathcal{A}^4}}}\left\{ {2{m^2}{{\left( {{a^2} - 2{r^2}} \right)}^2}{\mathcal{A}^4} + 4{m^2}{r^4}{{\left( {{a^2} + {r^2}} \right)}^2}{\mathcal{A}^2}{\mathcal{B}^2}} \right.
\nonumber \\
&+ {\left( {{a^2} + {r^2}} \right)^4}{\left( {{r^2}{\mathcal{A}^2}\mathcal{C} + {\varepsilon ^6}\sqrt {{a^2} + {r^2}} \mathcal{A}} \right)^2} + 2{\left( {{a^2} + {r^2}} \right)^2}\left[ {m{r^2}\mathcal{A}\mathcal{B}} \right. \
\nonumber \\
&{\left. { + \mathcal{C}\left( {{a^2} + {r^2}} \right)\left( {{r^2}{\mathcal{B}^2} + 12{a^2}{r^2}\left( {{r^2} - {\varepsilon ^2}} \right)\mathcal{A} - \mathcal{A}\mathcal{B}} \right)} \right]^2} + 2{\left( {{a^2} + {r^2}} \right)^2}
\nonumber \\
&\times \left[ {m{r^2}\mathcal{A}\mathcal{B} + {r^2}\left( {{a^2} + {r^2}} \right)\mathcal{C}{\mathcal{B}^2} - \left( {{a^2} + {r^2}} \right)\mathcal{A}\mathcal{C}\left( {{\varepsilon ^6} - 3{a^2}\left( {5{r^4} - } \right.} \right.} \right.
\nonumber \\
&{\left. {\left. {\left. {6{r^2}{\varepsilon ^2} + {\varepsilon ^4}} \right)} \right)} \right]^2} + {\left( {{a^2} + {r^2}} \right)^4}\left[ {{a^2}{\varepsilon ^6}{{\left( {{r^2} - {\varepsilon ^2}} \right)}^2}\left( {\sqrt {{a^2} + {r^2}} \left( {5{r^2} + {\varepsilon ^2}} \right)} \right.} \right.
\nonumber \\
&\left. { - 12m{r^2}} \right)\left. {{{\left. { - 9{a^4}{r^2}{{\left( {{r^2} - {\varepsilon ^2}} \right)}^4}\mathcal{C} + 2m{r^2}{\varepsilon ^{12}}} \right]}^2}} \right\},
\end{align}
while outside the gluing surface, it is
\begin{equation}
\label{eqa1-2}
{K^2}\left( {r > \varepsilon } \right) = \frac{{4{m^2}\left( {4{a^8} + 16{a^6}{r^2} + 29{a^4}{r^4} + 20{a^2}{r^6} + 12{r^8}} \right)}}{{{r^4}{{\left( {{a^2} + {r^2}} \right)}^5}}},
\end{equation}
where $ \mathcal{A} = {r^2}{\varepsilon ^6} - {a^2}{\left( {{r^2} - {\varepsilon ^2}} \right)^3}$, $\mathcal{B} = {\varepsilon ^6} - 3{a^2}{\left( {{r^2} - {\varepsilon ^2}} \right)^2}$, and $\mathcal{C} = 2m - \sqrt {{a^2} + {r^2}}$.

\setcounter{equation}{0}
\renewcommand\theequation{B.\arabic{equation}}
\section{The non-zero components of the stress-energy tensor of solution I}
\label{appB}
The non-zero components of the stress-energy tensor of Eq.~(\ref{eq10}) within the gluing surface are
\begin{align}
\label{eqb1-1}
T_t^t & = \frac{1}{{8\pi {{\left( {{a^2} + {r^2}} \right)}^{3/2}}{{\left[ {{r^2}{\varepsilon ^6} - {a^2}{{\left( {{r^2} - {\varepsilon ^2}} \right)}^3}} \right]}^2}}} \nonumber \\
& \times \left\{ {{a^2}{r^2}{\varepsilon ^6}\left[ {\mathcal{D} \left( {27{r^4}{\varepsilon ^2} - 3{r^2}{\varepsilon ^4} + {\varepsilon ^6} - 25{r^6}} \right) + m} \right.\left( {44{r^6}} \right.} \right.
\nonumber \\
&  \left. {\left. { - 42{r^4}{\varepsilon ^2} - 4{\varepsilon ^6}} \right)} \right] + 3{a^6}{\left( {{r^2} - {\varepsilon ^2}} \right)^4}\left( {7{r^2} - 2{\varepsilon ^2}} \right)\left( {\mathcal{D}  - 2m} \right)
\nonumber \\
&+ {a^4}\left( {{r^2} - {\varepsilon ^2}} \right)\left[ {\mathcal{D} \left( {21{r^{10}} - 69{r^8}{\varepsilon ^2} + 81{r^6}{\varepsilon ^4} - 64{r^4}{\varepsilon ^6} + 8{r^2}{\varepsilon ^8}} \right.} \right.
\nonumber \\
& \left. { - {\varepsilon ^{10}}} \right)\left. {\left. { - 2m\left( {18{r^{10}} - 57{r^8}{\varepsilon ^2} + 63{r^6}{\varepsilon ^4} - 53{r^4}{\varepsilon ^6} + 7{r^2}{\varepsilon ^8} - 2{\varepsilon ^{10}}} \right)} \right]} \right\},
\end{align}
\begin{align}
\label{eqb1-2}
T_r^r & = \frac{1}{{8\pi {{\left( {{a^2} + {r^2}} \right)}^{3/2}}{{\left[ {{r^2}{\varepsilon ^6} - {a^2}{{\left( {{r^2} - {\varepsilon ^2}} \right)}^3}} \right]}^2}}}
\nonumber \\
& \times \left\{ { - {a^2}{r^2}{\varepsilon ^6}\left[ {\mathcal{D}\left( {5{r^2} + {\varepsilon ^2}} \right){{\left( {{r^2} - {\varepsilon ^2}} \right)}^2} + m\left( {6{r^4}{\varepsilon ^2} - 4{r^6}} \right)} \right]} \right.
\nonumber \\
&+ 9{a^6}{r^2}{\left( {{r^2} - {\varepsilon ^2}} \right)^4}\left( {\mathcal{D} - 2m} \right) + {a^4}{\left( {{r^2} - {\varepsilon ^2}} \right)^2}\left[ \mathcal{D} \right.\left( {9{r^8} - 18{r^6}{\varepsilon ^2}} \right.
\nonumber \\
& \left. {\left. { \times \left. { + 9{r^4}{\varepsilon ^4} - 5{r^2}{\varepsilon ^6} - {\varepsilon ^8}} \right) + 6m{r^2}\left( {3{r^4}{\varepsilon ^2} + {\varepsilon ^6} - 2{r^6}} \right)} \right]} \right\},
\end{align}
\begin{align}
\label{eqb1-3}
T_\theta ^\theta & = T_\phi ^\phi  = \frac{1}{{{{8 \pi \left( {{a^2} + {r^2}} \right)}^{5/2}}{{\left[ {{r^2}{\varepsilon ^6} - {a^2}{{\left( {{r^2} - {\varepsilon ^2}} \right)}^3}} \right]}^2}}}
    \nonumber \\
& \times \left\{ {{a^2}{r^4}{\varepsilon ^6}\left[ {\mathcal{D} \left( {9{r^4}{\varepsilon ^2} + {\varepsilon ^6} - 10{r^6}} \right) + m\left( {16{r^6} - 12{r^4}{\varepsilon ^2} - {\varepsilon ^6}} \right)} \right]} \right.
\nonumber \\
&  + 3{a^8}{\left( {{r^2} - {\varepsilon ^2}} \right)^4}\left( {2{r^2} - {\varepsilon ^2}} \right)\left( {\mathcal{D}  - 2m} \right) + {a^6}\left( {{r^2} - {\varepsilon ^2}} \right)
\nonumber \\
&  \times \left[ {\mathcal{D} \left( {12{r^{10}} - 42{r^8}{\varepsilon ^2} + 54{r^6}{\varepsilon ^4} - 40{r^4}{\varepsilon ^6} + 5{r^2}{\varepsilon ^8} - {\varepsilon ^{10}}} \right)} \right.
\nonumber \\
& \left. {+ m\left( {55{r^8}{\varepsilon ^2} - 62{r^6}{\varepsilon ^4} + 46{r^4}{\varepsilon ^6} + {r^2}{\varepsilon ^8} + {\varepsilon ^{10}} - 17{r^{10}}} \right)} \right] + {a^4}{r^2}\left( {{r^2} - {\varepsilon ^2}} \right)
\nonumber \\
&  \times \left[ {\mathcal{D}} \right.\left( {6{r^{10}} - 21{r^8}{\varepsilon ^2} + 27{r^6}{\varepsilon ^4} - 35{r^4}{\varepsilon ^6} + {r^2}{\varepsilon ^8} - 2{\varepsilon ^{10}}} \right)
\nonumber \\
&\left. {\left. { + 2m\left( {14{r^8}{\varepsilon ^2} - 19{r^6}{\varepsilon ^4} + 28{r^4}{\varepsilon ^6} + 4{r^2}{\varepsilon ^8} + {\varepsilon ^{10}} - 4{r^{10}}} \right)} \right]} \right\},
\end{align}
whereas those outside the gluing surface  $\varepsilon$ can be expressed as
\begin{equation}
\label{eqb1-4}
T_t^t = T_r^r =  - \frac{{2{a^2}m}}{{{r^2}{\mathcal{D}^{3/2}}}}, \quad T_\theta ^\theta  = T_\phi ^\phi  = \frac{{3{a^2}m}}{{{\mathcal{D}^{5/2}}}}.
\end{equation}

For special but simple case $\varepsilon=a/2$,  the non-zero components of the stress energy tensor for Eq.~(\ref{eq10}) inside the gluing surface become
\begin{align}
\label{eqb1-5}
T_t^t & =\frac{1}{{8\pi}}\frac{1}{{{\mathcal{D}^{5}}{{\left( {{a^6} - 11{a^4}{r^2} + 48{a^2}{r^4} - 64{r^6}} \right)}^2}{{\left( { \mathcal{D}  - 2m} \right)}^2}}}
\nonumber \\
& \times  \left\{ {{a^{14}}\left( {136m - 23\mathcal{D} } \right) + 2{a^{12}}\left[ { \mathcal{D}  \left( {331{r^2} - 134{m^2}} \right) + 88{m^3} - 1988m{r^2}} \right]} \right.
\nonumber \\
&   + {a^{10}}{r^2}\left[ {\mathcal{D} \left( {8228{m^2} - 5855{r^2}} \right) - 5488{m^3} + 35488m{r^2}} \right]
   \nonumber \\
&   + 4{a^8}{r^4}\left[ { \mathcal{D} \left( {5753{r^2} - 19980{m^2}} \right) + 28m\left( {480{m^2} - 1261{r^2}} \right)} \right]
   \nonumber \\
&   - 16{a^6}{r^6}\left[ { \mathcal{D} \left( {1613{r^2} - 23044{m^2}} \right) + 2m\left( {7828{m^2} - 5197{r^2}} \right)} \right]
   \nonumber \\
&   - 64{a^4}{r^8}\left[ {\mathcal{D} \left( {11316{m^2} + 961{r^2}} \right) - 16m\left( {491{m^2} + 345{r^2}} \right)} \right]
   \nonumber \\
&   + 6144{a^2}{r^{10}}\left[ {\mathcal{D} \left( {8{m^2} + 13{r^2}} \right) - 12{m^3} - 79m{r^2}} \right]
   \nonumber \\
&  \left. { + 12288{r^{12}}\left[ { \mathcal{D} \left( {76{m^2} + 7{r^2}} \right) - 8\left( {6{m^3} + 5m{r^2}} \right)} \right]} \right\},
\end{align}
\begin{align}
\label{eqb1-6}
T_r^r & = \frac{{{\mathcal{D}^2}\left( {123{a^6}{r^2} - 1028{a^4}{r^4} + 1152{a^2}{r^6} + 2304{r^8} - {a^8}} \right)}}{{8{{\left( {{a^6} - 11{a^4}{r^2} + 48{a^2}{r^4} - 64{r^6}} \right)}^2}}}
\nonumber \\
& + \frac{{m{r^2}\left( {552{a^8}{r^2} - 3276{a^6}{r^4} + 7712{a^4}{r^6} - 3840{a^2}{r^8} - 6144{r^{10}} - 33{a^{10}}} \right)}}{{\pi {\mathcal{D}^3}{{\left( {{a^6} - 11{a^4}{r^2} + 48{a^2}{r^4} - 64{r^6}} \right)}^2}}},
\end{align}
\begin{align}
\label{eqb1-7}
T_\theta ^\theta & = T_\phi ^\phi  = \frac{{  288{a^8}{r^2} - 11{a^{10}} - 2544{a^6}{r^4} + 11648{a^4}{r^6} - 27648{a^2}{r^8} + 24576{r^{10}}}}{{8\pi {{\left( {{a^6} - 11{a^4}{r^2} + 48{a^2}{r^4} - 64{r^6}} \right)}^2}}}
\nonumber \\
&+ \frac{{m\left( {23{a^{14}} - 578{a^{12}}{r^2} + 4631{a^{10}}{r^4} - 18432{a^8}{r^6}} \right)}}{{8\pi {\mathcal{D}^5}{{\left( {{a^6} - 11{a^4}{r^2} + 48{a^2}{r^4} - 64{r^6}} \right)}^2}}}
\nonumber \\
&+ \frac{{m\left( {31168{a^6}{r^8} + 8704{a^4}{r^{10}} - 32768{a^2}{r^{12}} - 32768{r^{14}}} \right)}}{{8\pi {\mathcal{D}^5}{{\left( {{a^6} - 11{a^4}{r^2} + 48{a^2}{r^4} - 64{r^6}} \right)}^2}}},
\end{align}
where $\mathcal{D} = \sqrt {{a^2} + {r^2}}$.
\section{The Kretschmann scalar curvature of solution II}

\setcounter{equation}{0}
\renewcommand\theequation{C.\arabic{equation}}
\label{appD}
The Kretschmann scalar curvature of Eq.~(\ref{eq17}) within the gluing surface is
\begin{align}
\label{eqd1-1}
{K^2}\left( {\tau  \leq \varepsilon } \right) & = \frac{1}{{4{{\left( {{\varepsilon ^6}{\tau ^2} + l{\mathcal{X}^3}} \right)}^8}}}\left\{ {192{m^2}{\tau ^4}{\varepsilon ^{48}} + {l^3}{\mathcal{X}^5}{\varepsilon ^{30}}\left[ {4m\left( {{\varepsilon ^8} + 53{\tau ^2}{\varepsilon ^6}} \right.} \right.} \right.
\nonumber \\
&\left. { + 1758{\tau ^4}{\varepsilon ^4} + 1820{\tau ^6}{\varepsilon ^2} - 1688{\tau ^8}} \right) - 216{m^2}\left( {{\varepsilon ^6} + 11{\tau ^2}{\varepsilon ^4} + 84{\tau ^4}{\varepsilon ^2}} \right.
\nonumber \\
&\left. { - 60{\tau ^6}} \right)\left. { + 8{\tau ^2}\left( {136{\tau ^8} + 7{\varepsilon ^8} - 55{\tau ^2}{\varepsilon ^6} - 171{\tau ^4}{\varepsilon ^4} - 160{\tau ^6}{\varepsilon ^2}} \right)} \right]
\nonumber \\
& + {l^2}{\mathcal{X}^2}{\varepsilon ^{36}}\left[ {36{m^2}} \right.\left( {212{\tau ^8} + {\varepsilon ^8} + 26{\tau ^2}{\varepsilon ^6} + 277{\tau ^4}{\varepsilon ^4} - 372{\tau ^6}{\varepsilon ^2}} \right)
\nonumber \\
&- 16m{\tau ^2}\left( {230{\tau ^8}} \right. + 166{\tau ^2}{\varepsilon ^6}\left. { + 2{\varepsilon ^8} + 45{\tau ^4}{\varepsilon ^4} - 119{\tau ^6}{\varepsilon ^2}} \right)
\nonumber \\
&\left. { + 24{\tau ^4}\left( {34{\tau ^8} + 5{\varepsilon ^8} + 10{\tau ^2}{\varepsilon ^6} + 5{\tau ^4}{\varepsilon ^4}} \right)} \right] + l\left[ {288m{\varepsilon ^{24}}{\tau ^4}{\varepsilon ^{20}}\left( {{\varepsilon ^4} - {\tau ^4}} \right)} \right.
\nonumber \\
&\left. { - 96{m^2}{\tau ^2}{\varepsilon ^{42}}\left( {14{\tau ^6} + {\varepsilon ^6} + 24{\tau ^2}{\varepsilon ^4} - 39{\tau ^4}{\varepsilon ^2}} \right)} \right] + {l^4}{\mathcal{X}^8}{\varepsilon ^{24}}
\nonumber \\
&\left[ {324{m^2}\left( {36{\tau ^4} + {\varepsilon ^4} + 4{\tau ^2}{\varepsilon ^2}} \right) - 12m\left( {688{\tau ^6} + {\varepsilon ^6} + 26{\tau ^2}{\varepsilon ^4}} \right.} \right.
\nonumber \\
&\left. { + 392{\tau ^4}{\varepsilon ^2}} \right)\left. {\left. { + \left( {1568{\tau ^8} + 17{\varepsilon ^8} - 32{\tau ^2}{\varepsilon ^6} + 456{\tau ^4}{\varepsilon ^4} + 1312{\tau ^6}{\varepsilon ^2}} \right)} \right]} \right\},
\end{align}
while for the region outside the gluing surface, it simply becomes
\begin{equation}
\label{eqd1-2}
{K^2}\left( {\tau > \varepsilon } \right)  = \frac{{48{m^2}}}{{{\tau ^{12}}}},
\end{equation}
where $\mathcal{X}={\varepsilon ^2} - {\tau ^2}$.


\end{document}